\documentclass[fleqn,usenatbib]{mnras}

\usepackage{newtxtext,newtxmath}

\usepackage[T1]{fontenc}
\usepackage{ae,aecompl}

\usepackage{graphicx}	
\usepackage{amsmath}	
\usepackage{psfrag}
\usepackage{bm}
\usepackage[normalem]{ulem}

\def\V{\mathcal{V}}
\def\u{{\bf U}} 
\def\kv{{\mathbf k}}
\def\cl{{C_{\ell}}}
\newcommand{\vtha}{\mathbf{\theta}_1}

\title[Demonstrating the TGE]{ Demonstrating the Tapered Gridded Estimator (TGE) for the
 Cosmological HI 21-cm Power Spectrum using $150 \, {\rm MHz}$  GMRT observations}

\author[Pal S. et al.]{
Srijita Pal,$^{1}$\thanks{E-mail: srijitapal.phy@gmail.com}
Somnath Bharadwaj,$^{1}$\thanks{E-mail: somnathbharadwaj@gmail.com} 
Abhik Ghosh,$^{2}$
Samir Choudhuri$^{3}$
\\
$^{1}$ Department of Physics and Centre for Theoretical Studies, IIT Kharagpur, Kharagpur 721 302, India\\
$^{2}$ Department of Physics, Banwarilal Bhalotia College, GT Rd, Ushagram, Asansol, West Bengal 713303, India\\
$^{3}$ Astronomy Unit, Queen Mary University of London, Mile End Road, London E1 4NS, United Kingdom
}

\date{Accepted XXX. Received YYY; in original form ZZZ}

\pubyear{2020}

\begin{document}
\label{firstpage}
\pagerange{\pageref{firstpage}--\pageref{lastpage}}
\maketitle

\begin{abstract}
We apply the Tapered Gridded Estimator (TGE) for estimating the cosmological 21-cm power spectrum from $150 \, {\rm MHz}$ GMRT observations which corresponds to the neutral hydrogen (HI) at redshift $z = 8.28$. Here TGE is used to measure the Multi-frequency Angular Power Spectrum (MAPS) $C_{\ell}(\Delta\nu)$ first, from which we estimate the 21-cm  power spectrum  $P(k_{\perp},k_{\parallel})$. The data here are much too small for a detection, and the aim is to demonstrate the capabilities of the  estimator. We find that the estimated power spectrum is consistent with the expected foreground and noise behaviour. This demonstrates that this estimator correctly estimates the noise bias and subtracts this out to yield an unbiased estimate of the power spectrum. More than $47\%$ of the frequency channels had to be discarded from the data owing to radio-frequency interference, however the estimated power spectrum does not show any artifacts due to missing channels. Finally, we show that it is possible to suppress the foreground  contribution  by tapering the sky response at large angular separations from the phase center. We combine the k modes within a rectangular region in the `EoR window' to obtain the spherically binned averaged dimensionless power spectra $\Delta^{2}(k)$ along with the statistical error $\sigma$ associated with the measured $\Delta^{2}(k)$. The lowest $k$-bin yields $\Delta^{2}(k)=(61.47)^{2}\,{\rm K}^{2}$ at $k=1.59\,\textrm{Mpc}^{-1}$, with $\sigma=(27.40)^{2} \, {\rm K}^{2}$. We obtain a $2 \, \sigma$ upper limit of $(72.66)^{2}\,\textrm{K}^{2}$ on the mean squared HI 21-cm brightness temperature fluctuations at $k=1.59\,\textrm{Mpc}^{-1}$.
\end{abstract}


\section{Introduction}
\label{s1}
Measurements of the cosmological HI 21-cm power spectrum can be used to probe the large
scale distribution of neutral hydrogen (HI) across a large redshift range from the Dark Ages
to the Post-Reionization Era
 \citep{BA5,furla06,morales10,prichard12,mellema13}. Several radio interferometers such as
 the Giant Meterwave Radio Telescope (GMRT\footnote{\url{http://www.gmrt.ncra.tifr.res.in/}}; 
 \citealt{swarup91}), the Low Frequency Array (LOFAR\footnote{\url{http://www.lofar.org/}}, \citealt{haarlem}),  
the Murchison Wide-field Array (MWA\footnote{\url{http://www.mwatelescope.org}} \citealt{tingay13}), 
and the Donald C. Backer Precision Array to Probe the Epoch of Reionization (PAPER\footnote{\url{http://eor.berkeley.edu/}}, \citealt{parsons10})
 have carried out observations to measure the 21-cm power spectrum from the Epoch of Reionization (EoR). Despite  of the ongoing efforts, only a few upper limits on the power spectrum amplitudes have
been reported in the literature till date (e.g. GMRT: \cite{paciga11,paciga13}; LOFAR: \cite{yata13,patil17,Gehlot19,mertens20,Mondal20a}; MWA: \cite{dillon14,Jacobs16,Li19,Barry19,Trott20}; PAPER: \cite{Cheng18,Kolopanis19}). A few more upcoming telescopes such as the Hydrogen Epoch of Reionization Array (HERA\footnote{\url{http://reionization.org/}}; \citep{DB17}) and the Square Kilometer Array (SKA\footnote{\url{http://www.skatelescope.org/}}; \citep{Koo15})  also aim to measure
the EoR 21-cm power spectrum with improved sensitivity. 

The primary challenge for detecting  the redshifted 21-cm signal are the foregrounds
which include extra-galactic point sources (EPS), the diffuse Galactic synchrotron emission (DGSE), the free-free emission from our Galaxy and external galaxies \citep{shaver99,dmat1,santos05, ali,Bern09,paciga11,ghosh3,Ia13,samir17a}. The foregrounds are three to four orders of magnitude larger than the expected 21-cm signal. A variety of techniques have been proposed to overcome this issue. Among these, 
`foreground removal'  proposes  to subtract out a foreground model from the data  and use the residual data to detect the 21-cm power spectrum \citep{jelic08,bowman09, paciga11, chapman12, trott1,trott16}. Recently, a novel foreground removal method based on Gaussian Process Regression has been used to model and remove foregrounds from LOFAR \citep{mertens18, mertens20} and HERA data \citep{Ghosh20}. Further, the foregrounds are predicted to be  primarily confined to a wedge shaped region in the $(k_{\perp}, \, k_{\parallel})$ plane. The  `foreground avoidance' technique proposes to use the region outside this so called  Foreground Wedge to estimate the 21-cm power spectrum   \citep{adatta10,vedantham12,thyag13,pober13,pober14,liu14a,liu14b,dillon14,dillon15,zali15}. Bright sources located at a considerable angular distance from the phase center (wide field foregrounds) are particularly important for measuring the 21-cm power spectrum. It is extremely challenging to model and subtract out such sources due to ionospheric fluctuations   and 
lack of knowledge of the primary beam (PB) pattern far away from the phase center. The contribution from such sources manifest themselves as oscillatory  frequency structures \citep{ghosh1,ghosh2}. Several studies (e.g. \citealt{thyag15,pober16}) have shown that such sources  contaminates the higher $k_{\parallel}$ modes which are relevant for measuring the 21-cm power spectrum. 
The polarization leakage  is also expected to increase with distance from the phase center \citep{Asad15, Asad18}.

Various   estimators have been proposed for the 21-cm power spectrum.  Image-based estimators (e.g. \citealt{paciga13}) have the disadvantage that deconvolution errors which arises during image reconstruction may affect the estimated power spectrum.  Techniques like the Optimal Mapmaking Formalism \citep{morales09} avoid this deconvolution error during imaging. This problem of deconvolution does not arise if the power spectrum is directly estimated from  the measured visibilities
 \citep{morales05,mcquinn06,pen09,liu12,parsons12,liu14a,liu14b,dillon15,trott16}. 
\citet{liu16} have accounted for  the sky curvature by using the spherical Fourier-Bessel basis to estimate the power spectrum. The noise bias arising from the  noise contribution present in the measured visibilities (or the image) is also an issue for power spectrum estimation.  For example, \citet{zali15} have avoided this by dividing the data 
 into even and odd LST bins  and have  correlated these  to estimate the power spectrum. However, the full signal available in the data is not used in such an approach. In an alternative  approach, several 21-cm power spectrum estimators  have been proposed  \citep{Shaw14,Shaw15,Eastwood19,Patwa19}  for  drift scan observations. Another alternative approach to detect the 21-cm signal \citep{Thyagarajan18,Thyagarajan20} uses the fact that the interferometric bispectrum phase is immune to antenna-based calibration errors.  
 
The Tapered Gridded Estimator (TGE) is a novel visibility based power spectrum estimator presented initially  for the angular power spectrum $C_{\ell}$ in 2D \citep{samir14,samir16} and subsequently for the  3D power spectrum $\textrm{P}(k_{\perp},k_{\parallel})$ \citep{samir17}. The TGE suppresses the contribution from sources far away from the phase center by tapering the sky response with a tapering window function which falls off faster than the PB. Further, TGE 
also internally estimates and subtracts out the  noise bias,  and provides  an unbiased estimate of the power spectrum. 
TGE has the added advantage that it works with the gridded  visibility data which makes the estimator 
  computationally fast, an important factor for  future telescopes like SKA-I which are expected to produce large amounts of data. Several studies have used the 2D TGE  to measure  the angular power spectrum $C_{\ell}$ of the DGSE using GMRT data at $150 \, {\rm MHz}$   \citep{samir17a,samir20} and also at $325 \, {\rm MHz}$  \citep{Cha1,Cha2, M20}. \citet{Saha19} have used the 2D TGE to measure $C_{\ell}$   of the fluctuations in the synchrotron emission from  the Kepler supernova remnant. \citet{ITGE} have developed an Image-based Tapered Gridded Estimator (ITGE) which was used to measure $C_{\ell}$  of the HI 21-cm emission from the ISM in different parts of of an  external galaxy.  These studies clearly establish the 2D TGE as an efficient and reliable estimator for the angular power spectrum $C_{\ell}$, and also demonstrate its ability to suppress the contribution from sources which are far away from the phase center.    
  
  The Multi-frequency Angular Power Spectrum $C_{\ell}(\nu_a, \nu_b)$  (MAPS; \citealt{KD07,Mondal19}) characterizes the joint angular and frequency dependence of the sky signal. It is relatively straightforward to generalise the 2D TGE for $C_{\ell}$ to a 3D TGE for $C_{\ell}(\nu_a, \nu_b)$ . In  our previous work (\citep{Bh18}, hereafter, Paper I) we present a TGE estimator for MAPS, and use this to propose a new technique to estimate the 3D power spectrum $P(k_{\perp}, k_{\parallel})$    of the cosmological 21-cm signal.  This  has been validated using simulations in Paper I.  While this retains all the aforementioned advantages of the 2D TGE, it has some  additional advantages arising from the fact that we first evaluate the binned MAPS  $C_{\ell}(\Delta \nu)$ where $\Delta \nu= \mid \nu_a-\nu_b \mid$, and then use this to estimate  $P(k_{\perp}, k_{\parallel})$ through a Fourier transform with respect to $\Delta \nu$. This is in contrast to the usual approach (e.g. \citealt{Morales04})
  where the individual visibilities are first Fourier transformed along frequency and then correlated to estimate  $P(k_{\perp}, k_{\parallel})$.  Considering the advantages of our new approach, first it is not necessary to introduce a frequency filter to ensure continuity at the edge of the frequency band. Second, it is computationally inexpensive to implement a maximum likelihood estimator for the Fourier transform  \citep{trott16} as the  data volume is considerably reduced if we consider the binned $C_{\ell}(\Delta \nu)$ instead of the individual visibilities. Finally, and most important, our new approach is relatively unaffected by missing frequency channels due to flagging. The simulations in Paper I demonstrate that our  new estimator is able to accurately recover the input model power spectrum even in a situation when $80 \, \%$ randomly chosen frequency channels in the data  are flagged. 
  
In this paper we demonstrate the capabilities of the new estimator by applying it to  GMRT observations centred at $153 \, {\rm  MHz}$  which corresponds to the $21$-cm signal from $z = 8.28$. This  data has been analysed earlier to 
characterize the statistical properties of the foregrounds \citep{ghosh3}. This is a relatively short observation where the total observation time is $11$ hrs. Further, more than $47 \%$ of the data had to be flagged to avoid Radio Frequency Interference (RFI) and other systematic errors. In this paper we have applied the TGE to  this data to estimate  the MAPS $C_{\ell}(\Delta \nu)$  and the 3D power spectrum $P(k_{\perp}, k_{\parallel})$.

This paper has been arranged as follows. Section 2 gives a brief description of the GMRT observation and some details of the initial reduction of the data. In Section 3 we briefly summarize the methodology used to estimate $C_{\ell}(\Delta \nu)$  and $P(k_{\perp}, k_{\parallel})$, and in Section 4  we  validate our estimator using  simulations which have exactly the same baseline distribution and flagging as the actual data. 
In Section 5 we present the results for $C_{\ell}(\Delta \nu)$  and $P(k_{\perp}, k_{\parallel})$.
We also identify a rectangular region outside the foreground wedge and combine all the $(k_{\perp}, k_{\parallel})$ modes within it to obtain upper limits on the 21-cm brightness temperature fluctuations. We present summary and conclusions in 
 Section 6. 
 
 We have used $\Lambda$CDM cosmology and Planck+WMAP9 \citep{PLANCK16b} best fit cosmological parameters throughout this paper unless mentioned otherwise.

\section{Observation and Data Analysis}
\label{s2}

The GMRT has a hybrid configuration \citep{swarup91}, where 14 antennas each with 45 m diameter are randomly distributed in a Central Square which is approximately $1.1 \, {\rm km} \times 1.1 \, {\rm km}$ in extent. The rest of the 16 antennas lie along three $\sim 14$ km long arms in an approximately `Y' shaped configuration. The hybrid configuration of the GMRT, with the shortest and largest baseline separation of approximately $60 \, {\rm m}$  and $26  \, {\rm km}$ respectively, gives reasonably good sensitivity to probe both compact and extended sources. 

The target field (FIELD I of \citealt{ghosh3}) was  observed in GTAC (GMRT Time Allocation Committee) cycle 15 in January 2008. In this section, we present a brief summary of the observational parameters and initial processing of the data relevant to this paper. For further information and a more detailed description, the reader is referred to (\cite{ghosh3}, Section $2$). The relevant parameters for this observation are summarized in Table \ref{t_1}. The visibilities were recorded for two circular polarizations (RR and LL) with $128$ frequency channels covering a bandwidth of $8$ MHz and an integration time of $16$s. The field contains relatively few bright sources ($\geq 0.3$ Jy) in the $1400$ MHz NRAO VLA Sky Survey (NVSS) and have relatively low sky temperature ($\sim40$ K) with no significant structure visible at an angular resolution of $\sim 0.85^{\circ}$ in the $408$ MHz Haslam map. The flux density of the brightest source in the field is $905$ mJy/beam at 153 MHz. The field is situated at a high galactic latitude and was observed at night time to minimise the RFI from man made sources. Further, the ionosphere is considerably more stable at night, and the phase errors that can vary significantly with time due to ionospheric scintillation is expected to be less severe.

\begin{table}
\caption{Observation summary} 
\label{t_1}
\begin{tabular}{|l|c|}
\hline
\hline
Central Frequency $(\nu_c)$ & $153$ MHz  \\
\hline
Channel width $(\Delta\nu_c)$ & $62.5$ kHz \\
\hline
Bandwidth $(B_{bw})$ & $8.00$ MHz \\
\hline
Total observation time &  $11 $ hrs \\
\hline
Target field  $(\alpha,\delta)_{2000}$ &  ($05^h30^m00^s$,\\ 
 & $+60^{\circ}00^{'}00^{''}$) \\
 \hline
 Galactic coordinates $(l,b)$ & $151.80^{\circ}, 13.89^{\circ}$ \\
\hline
Off source noise & $1.3$ mJy/Beam \\
\hline
Flux density (max., min.) &  ($905$ mJy/Beam, \\
& $-14$ mJy/Beam) \\
\hline
Synthesized beam & $21^{''} \times 18^{''}$ , PA = $61^{\circ}$\\
\hline
Comoving distance at $153$ MHz $(r)$ & $9231$ Mpc \\
\hline
$r^{\prime}$ at $153$ MHz $(dr/d\nu)$ & $16.99$ Mpc/MHz \\
\hline
\hline
\end{tabular}
\end{table}

\begin{figure}
\begin{center}
\psfrag{channel}[rt][lb][1.5]{$\textrm{channel no.}$}
\psfrag{ 20}[c][b]{$20$}\psfrag{ 30}[c][b]{$30$}\psfrag{ 40}[c][b]{$40$}\psfrag{ 50}[c][b]{$50$}\psfrag{ 60}[c][b]{$60$}\psfrag{ 70}[c][b]{$70$}\psfrag{ 80}[c][b]{$80$}
\psfrag{RR}[ct][rb][0.8]{$\textrm{RR}$}\psfrag{LL}[t][b][0.8]{$\textrm{LL}$}
\psfrag{U1}[ct][rb][0.8]{$ $}\psfrag{F1}[ct][rb][0.8]{$\approx91\%$}
\psfrag{U2}[ct][rb][0.8]{$ $}\psfrag{F2}[ct][rb][0.8]{$\approx92\%$}
\includegraphics[width=85mm]{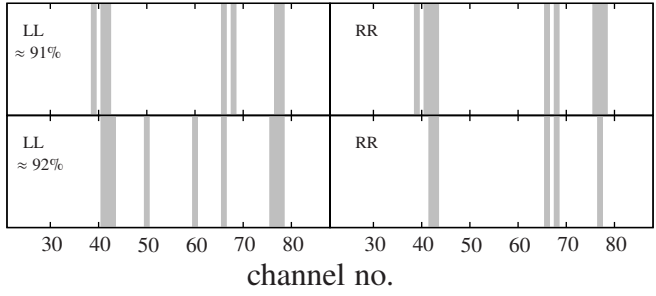}\vspace{8mm}
\caption{Unflagged frequency channels (shown in grey) for two randomly chosen baselines  $U=82 \, \lambda$ (upper panels) and $2688 \, \lambda$ (lower panels) for which   respectively $91\%$ and $92\%$ of  the channels are flagged. Stokes LL and RR are shown in the left and right panels respectively.}
\label{flag}
\end{center}
\end{figure}

The flagging and calibration of the data were done using the software called {\scriptsize FLAGCAL} \citep{jayanti}. At low frequencies RFI is a major challenge limiting the sensitivity of the array. The software {\scriptsize FLAGCAL} identifies and removes bad visibilities by requiring that good visibilities be continuous in time and frequency, and then using known flux and phase calibrators computes calibration solutions and interpolates them onto the target fields using spherical linear interpolation (slerp).  The RFI problem is particularly severe for the GMRT at the low frequency bands. Figure \ref{flag} shows the data across the frequency for two visibility records, chosen randomly, to highlight the heavy flagging (more than $90 \%$) for these two baselines. Note, on average the flagging fraction across all the baselines  within $3000\lambda$ is around $47 \%$. In this observation, the gain solutions were calculated using a known flux calibrator (3C147), observed at the beginning of the observation, and phase calibrator (3C147), observed every half an hour during the whole observation run. These gain solutions were then interpolated and applied on the target field centred at $\alpha_{2000} = 05^h30^m00^s$, and $\delta_{2000} = +60^{\circ}00^{'}00^{''}$. Flagcal does a two point interpolation in time. There are two options available, linear interpolation and  spherical linear interpolation (slerp). We used the default option, spherical linear interpolation. Subsequently, {\scriptsize AIPS} task {\scriptsize IMAGR} was used to image the field which was then self-calibrated (three rounds of phase calibration followed by one round for both amplitude and phase) using the bright sources present in the field with a solution time interval of 5,3,2 and 5 min for the successive self-calibration loops. The final gain table was applied to all the 128 frequency channels centred at $153$ MHz.

The discrete point sources within a field of view (FoV)  of $4.0^{\circ} \times 4.0^{\circ}$
dominate the 150 MHz radio sky at the angular scales probed in our observations.
These are mainly associated with active galactic nuclei (AGN). It creates a major problem for detecting the redshifted 21-cm signal at the arc-minute angular scales probed in our observation. Simulations \citep{bowman09, Liu09} also suggest that point sources should be subtracted down to 10-100 mJy level in order to detect the EoR signal. Here, we have subtracted all the point sources using the 
{\scriptsize AIPS} task {\scriptsize UVSUB} from the entire FoV (mostly within the main lobe of the PB) above a threshold flux level of 7 times the r.m.s. noise ($\sigma=1.3$ mJy/Beam).

It is expected that at this stage most of the genuine sources above a flux level of $\sim 9$ mJy have been removed from the uv data. 
A visual inspection of the image of the field shows that most of the imaging artifacts are in a few regions in the image, typically close to the bright sources and were not modelled using the clean components. After source subtraction the  
resulting image had a maximum flux density of $21$ mJy/Beam and minimum flux density of $-14$ mJy/Beam. The visibilities before and after source subtraction are used in the rest of the analysis. Further, we have used $88$ frequency channels from the central part of the frequency band, two polarizations (RR and LL) and a baseline range of $ 70 \lambda \leq \lvert {\bf U}_{i} \rvert \leq 3000 \lambda$ in the subsequent analysis.\\

\section{Estimating  MAPS and the 3D Power Spectrum}
\label{s3}

The MAPS $C_{\ell}(\nu_a, \nu_b)$ quantifies  the  statistics  of the  sky signal as a joint  function of angular multipole and frequency. 
This second order statistic completely quantifies the statistical properties of the sky signal 
under the assumption that the brightness temperature fluctuations are generated as a result of a Gaussian random process that is statistically homogeneous and isotropic on the sky.  Considering a  particular frequency $\nu$, 
the brightness temperature fluctuations across the sky $\delta T_{\rm b} (\hat{\bm{n}},\,\nu)$, can be decomposed in terms of spherical harmonics $Y_{\ell}^{\rm m}(\hat{\bm{n}})$ as,
\begin{equation}
\delta T_{\rm b} (\hat{\bm{n}},\,\nu)=\sum_{\ell,m} a_{\ell {\rm m}} (\nu) \,
Y_{\ell}^{\rm m}(\hat{\bm{n}}).
\label{eq:alm}
\end{equation}
The MAPS is then defined as
\begin{equation}
C_{\ell}(\nu_a, \nu_b) = \big\langle a_{\ell {\rm m}} (\nu_a)\, a^*_{\ell
  {\rm m}} (\nu_b) \big\rangle\,
\label{eq:cl}
\end{equation}
where the angular brackets $\langle  ... \rangle$ denote an ensemble average over different statistically independent realizations of the Gaussian random field  $\delta T_{\rm b} (\hat{\bm{n}},\,\nu)$.
In Paper I we have established a TGE to  estimate  MAPS directly from  the visibility data $\mathcal{V}_{i}(\nu_{a})$ which is the fundamental quantity measured in radio interferometric observations. Here $\nu_{a}$ refers to the different frequency channels  with $a=1,2,3,...,N_c$ where $N_c$ refers to the total  number of frequency channels which span a  frequency bandwidth $B_{bw}$.
 The observed visibilities are convolved with a function $\tilde{w}(\u)$ and gridded on a rectangular grid in the uv-plane using,
\begin{equation}
\V_{cg}(\nu_{a}) = \sum_{i}\tilde{w}(\u_g-\u_i) \, \V_i(\nu_{a}) \,F_i(\nu_a)
\label{eq:a1}
\end{equation}
where $\textbf{U}_{i}$ is the baseline corresponding to the $i$-th visibility,
$\V_{cg}$ refers to the convolved visibility at grid point $g$, $\u_{g}$ is the baseline corresponding to this grid point and  $F_i(\nu_a)$ incorporates  the flagging information of the data corresponding to the frequency $\nu_a$. $F_i(\nu_a)$ has a value `$0$' if the data at a given baseline and frequency is flagged and `$1$' otherwise. Note that the baselines $\u_i$ here are defined at a fixed reference frequency $\nu_c$, and they do not change as we vary the frequency channel $\nu_a$. 

The  sky response of the convolved  visibility $\V_{cg}$  is tapered by the window function ${\cal W}(\theta)$, which is the Fourier transform of $\tilde{w}(\u)$. The main lobe of the PB pattern of GMRT (or any other telescope with a circular aperture) can be approximated by a Gaussian $A(\theta)=e^{-\theta^{2}/\theta^{2}_{0}}$ where  $\theta_{0}$ is approximately $0.6$ times the full width half maxima ($\theta_{FWHM}$) of the Gaussian \citep{BS01, samir14} which has    a value 
$\theta_{0}=95^{'}$ that  corresponds to $\theta_{FWHM}=158^{'}$ at $\nu_{c}=153$ MHz for the GMRT.
Here  we have used 
\begin{equation}
{\cal W}(\theta)=e^{-\theta^{2}/\theta^{2}_{w}}
\label{eq:win1}
\end{equation}
 with $\theta_{w}=f\theta_{0}$ where $f$ is a parameter whose value  can be suitably chosen. The effect of tapering is enhanced if the value of $f$ is reduced. 
 A value $f <1$ would suppress  the sky response from the outer regions and the side-lobes of the PB pattern, whereas a large value $f >1$ would imply very little tapering. 
 
We define the TGE as
\begin{eqnarray}
{\hat E}_g(\nu_a,\nu_b) &=& M_g^{-1}(\nu_a,\nu_b) \, 
{\mathcal Re} \Big[\V_{cg}(\nu_a) \,  \V_{cg}^{*}(\nu_b) \, \nonumber \\
&& - \, \sum_i F_i(\nu_a)F_i(\nu_b)  \mid
\tilde{w}(\u_g-\u_i) \mid^2   \V_i(\nu_a) \V_i^{*}(\nu_b)  \Big] \, \nonumber \\
\label{eq:a4}
\end{eqnarray}

where ${\mathcal Re}[Z]$ refers to the real part of the expression $Z$. We note that eq.~(\ref{eq:a4}) is slightly different  from the TGE estimator which was defined and validated in Paper I. The second term in the brackets has been introduced to subtract out the positive definite noise bias which arises due to the noise contribution present in each visibility. Here we assume that the noise in different visibilities is uncorrelated and only the self-correlation of a visibility ({\it i.e.} same baseline, frequency channel and timestamp) contributes to the noise bias. Thus only the self-correlation of a visibility ({\it i.e.} same baseline, frequency channel and timestamp) contributes to the noise bias. So it is adequate if the self-correlation term is subtracted from visibility correlations only when $\nu_a = \nu_b$ {\it i.e.} 
$\Delta \nu =\mid \nu_{a}-\nu_{b}\mid=0 $, and  this is what was implemented and validated in the estimator proposed in Paper I.  However, in the presence of foregrounds  we find that this causes an abrupt dip in the estimated MAPS at $\Delta \nu=0$. This dip is not particularly prominent for the present data where it is comparable to the uncertainty arising from  the system noise. However, this dip introduces a negative bias in the estimated 3D power spectrum, and for the present data this becomes noticeable when  we consider the spherically binned power spectrum $P(k)$ (discussed later). The estimator presented in eq.~(\ref{eq:a4}) overcomes this 
problem by subtracting out the self-correlation for all combinations of $\nu_a$ and   $\nu_b$. The dip mentioned above, and the resulting negative bias in the estimated power spectrum,  become particularly prominent for more sensitive data.  We plan to present a detailed comparison of the earlier  estimator with the modified one in a future paper considering  more sensitive data.

 $M_g(\nu_a,\nu_b)$ in eq.(~\ref{eq:a4})  is a   normalisation factor whose values is determined  using simulations. For this we simulate observations having the same baseline and frequency coverage and also the same flagging as the actual data analyzed here. The simulated sky signal corresponds to a  Gaussian random  field with an unit multi-frequency angular power spectrum (UMAPS) for which $ C_{\ell}(\nu_a, \nu_b)=1$.  We use the simulated visibilities $ [\V_i(\nu_a)]_{\rm UMAPS}$ to determine the normalization factors through
 \begin{eqnarray}
M_g(\nu_a,\nu_b) \, &=& \, 
{\mathcal Re} \Big[ \V_{cg}(\nu_a) \,  \V_{cg}^{*}(\nu_b) - \, \nonumber \\
&& \sum_i F_i(\nu_a) F_i(\nu_b) \mid
\tilde{w}(\u_g-\u_i) \mid^2  
  \V_i(\nu_a) \V^{*}_i(\nu_b)  \Big]_{\rm UMAPS} \, \nonumber \\
\label{eq:a4a}
\end{eqnarray}

We have averaged over multiple realisations of UMAPS  
in order to reduce the statistical uncertainty in the estimated  values of $M_g(\nu_a,\nu_b)$.

The estimator  ${\hat E}_g(\nu_a,\nu_b)$ provides an unbiased estimate of the MAPS  $\langle {\hat E}_g(\nu_a,\nu_b) \rangle =C_{\ell_g}(\nu_a,\nu_b)$  at the grid point $\u_g$ which corresponds to an angular multipole  $\ell_g=2\,\pi\,\mid \u_g \mid$. 
In order to increase the signal to noise ratio, we  bin the entire $\ell$ range into bins of equal logarithmic interval. 
Considering a bin labelled `$q$', we define 
the bin averaged Tapered Gridded Estimator as 
\begin{equation}
{\hat E}_G[q](\nu_a,\nu_b) = \frac{\sum_g w_g  {\hat E}_g(\nu_a,\nu_b)}
{\sum_g w_g } \,.
\label{eq:a6}
\end{equation}
where the sum is over all the grid points $\u_g$ included in the  particular bin   and  the $w_g$'s are weights corresponding to the different grid points. In this paper we have used  
$w_g=M_g(\nu_a,\nu_b)$  where the weight is proportional to the baseline sampling of the particular grid point. 
The ensemble average of ${\hat E}_G[q](\nu_a,\nu_b)$ gives an unbiased estimate of $\bar{C}_{\bar{\ell}_q}(\nu_a,\nu_b)$ which is the bin averaged multi-frequency angular power spectrum   at the effective angular multipole $\bar{\ell}_q = \frac{ \sum_g w_g \ell_g}{ \sum_g w_g}$.  

We now discuss how the MAPS $C_{\ell}(\nu_a, \nu_b)$ can be used to estimate the 3D power spectrum $P(k_{\perp}, k_{\parallel})$ of the redshifted 21-cm brightness temperature fluctuations. Here 
the redshifted 21-cm signal  is assumed to be statistically
homogeneous (ergodic) along the line of sight comoving distance (e.g. \citealt{Mondal19}). 
Considering a small bandwidth we then  have
$\cl(\nu_a,\nu_b)=\cl(\Delta \nu)$ where $\Delta \nu = \mid \nu_b-\nu_a \mid$, 
{\it i.e.} the statistical properties only depend 
on the  frequency separation and not the individual frequencies. 
Under the flat sky approximation,  the power spectrum $P(k_{\perp}, k_{\parallel})$ 
is  the Fourier transform of
$\cl(\Delta \nu)$, and we have \citep{KD07}  
\begin{equation}
P(k_{\perp},\,k_{\parallel})= r^2\,r^{\prime} \int_{-\infty}^{\infty}  d (\Delta \nu) \,
  e^{-i  k_{\parallel} r^{\prime} \Delta  \nu}\, C_{\ell}(\Delta \nu)
\label{eq:cl_Pk}
\end{equation}
where $k_{\parallel}$ and $k_{\perp}=\ell/r$ are the components of $\kv$ respectively parallel and perpendicular to the line of sight, $r$ and $r^{\prime}=dr/d\nu$ are respectively the comoving distance and its derivative with respect to $\nu$, both evaluated at the reference frequency $\nu_{c}$. Here, throughout the analysis we have used $r=9231$ Mpc and $r^{\prime}=16.99$ Mpc/MHz at $\nu_{c}=153$ MHz.\\

We have estimated the $C_{\ell}(n \, \Delta\nu_{c})$ for  $-(N_{c}-1)\leq n\leq(N_{c}-1)$ discrete channel separations. 
The fact that $C_{\ell}(n\Delta\nu_{c})=C_{\ell}(-n\Delta\nu_{c})$ implies that  $C_{\ell}(n \, \Delta\nu_{c})$ is periodic in $n$ 
with a period of $2(N_c-1)\Delta\nu_c$. In Paper~I we have used the discrete Fourier transform
\begin{equation}
\bar{P}(k_{\perp},\,k_{\parallel m}) = (r^2\,r^{\prime} \, \Delta \nu_c) 
\sum_{n=-N_c+2}^{N_c-1}   \exp \left( -i  k_{\parallel m} r^{\prime} \, n  \, \Delta  
\nu_c \right) \, C_{\ell}(n \, \Delta \nu_c) .
\label{eq:b1}
\end{equation}
to obtain the 3D power spectrum from  the estimated bin  averaged $C_{\ell}(n \, \Delta\nu_{c})$.
 Eq. (\ref{eq:b1}) provides  an  estimate of the 3D power spectrum $\bar{P}(k_{\perp},\,k_{\parallel m})$ for 
  $k_{\parallel  m}=m \, [\pi/r^{\prime}_{\rm c} \, \Delta \nu_c (N_c-1)]$ where $-N_c+2\leq m \leq N_c-1$. 
 
Paper I has used simulations with a known input power spectrum to demonstrate that the TGE, in its old form, along with the discrete Fourier transform outlined above, is able to accurately recover the input power spectrum from the simulated visibilities even in the presence of heavy flagging. In this paper we incorporate  a further improvement by replacing the discrete Fourier transform in eq.~(\ref{eq:b1}) with a maximum likelihood estimator. The issue here is that  the $C_{\ell}(n \, \Delta \nu_c) $ estimated at each $n \, \Delta \nu_c$ contributes with equal weight  in eq.~(\ref{eq:b1}). We however note that in the absence of flagging there are  $N_c-n$ independent channel pairs corresponding to any particular  channel separation $n \Delta \nu_c$, implying that $C_{\ell}( n \, \Delta \nu_c)$ is better estimated for the smaller $n$ as compared to the larger $n$. It is desirable to incorporate  this  by introducing different  weights  for each  $C_{\ell}(n \, \Delta \nu_c) $ when estimating the 3D power spectrum. We adapt a new technique to estimate   $\bar{P}(k_{\perp},k_{\parallel m})$
 from  $C_{\ell}(n \, \Delta \nu_c) $  based on Maximum Likelihood Estimation  (MLE) which we describe as follows.\\

The inverse of eq. (\ref{eq:b1}) can be recast in the matrix notation as
\begin{equation}
    C_{\ell}(n \, \Delta\nu_{c})=  \sum_{m} \textbf{A}_{nm} \, \bar{P}(k_{\perp},k_{\parallel m}) + [\textrm{Noise}]_{n}
\label{eq:b2}
\end{equation}
where $n,\,m\, \epsilon\, [0,N_{c}-1]$. Here the estimated 
$C_{\ell}(n \, \Delta\nu_{c})$ is modelled as  the Fourier transform of the 3D power spectrum $\bar{P}(k_{\perp},k_{\parallel m})$  plus an additive noise $[\textrm{Noise}]_{n}$.  $\textbf{A}_{nm}$ here refers to the components of the  $N_{c} \times N_{c}$ Hermitian matrix $\textbf{A}$ corresponding to  the Fourier transform  coefficients.
 The maximum likelihood estimate of $\bar{P}(k_{\perp},k_{\parallel m})$ is  given by 
\begin{equation}
   \bar{P}(k_{\perp},k_{\parallel m}) =  \sum_n \{ [\textbf{A} ^{\dagger} \textbf{N}^{-1} \textbf{A}]^{-1} \textbf{A}^{\dagger}    \textbf{N}^{-1} \}_{mn}  C_{\ell}(n\Delta\nu_{c})
    \label{eq:ML}
\end{equation}
where $\textbf{N}$ is the noise covariance matrix and `$\dagger$' denotes  the Hermitian conjugate. 
Here we have used `noise-only' simulations to estimate $\textbf{N}$.  For these simulations  each measured visibility  is  assigned random Gaussian noise, the noise in the different visibilities is assumed to be uncorrelated. The simulated visibilities were used to estimate $C_{\ell}(n \,  \Delta \nu_c)$.  The $C_{\ell}(n \,  \Delta \nu_c)$ estimated from 
multiple statistically independent noise realizations were  used to estimate the noise  covariance matrix $\textbf{N}$.

 We have further binned $\bar{P}(k_{\perp},\,k_{\parallel m})$ in $k_{\parallel m}$ to obtain the bin averaged Cylindrical Power Spectrum $P(k_{\perp}, k_{\parallel})$ which we show in the subsequent analysis. Here we find it convenient to use  bins of equal linear spacing for $k_{\parallel}$.

\section{Simulation}
\label{sim}

\begin{figure}
\begin{center}
\psfrag{freq. seper.}[1.0]{$\Delta\nu\hspace{.2cm} [\textrm{MHz}]$}
\psfrag{n2172}[c][0.8]{$\textrm{NF}$}\psfrag{n4747}[c][0.8]{$\textrm{NF}$}\psfrag{n16426}[c][0.8]{$\textrm{NF}$}
\psfrag{2172}[c][l][0.8]{$\ell=2163$}\psfrag{4747}[c][l][0.8]{$\ell=4759$}\psfrag{16426}[c][l][0.8]{$\ell=16430$}
\psfrag{-5.000000e-09}[c][1.0]{$-0.5\hspace{.5cm}$}\psfrag{0.000000e+00}[r][1.0]{$ 0.0\hspace{.1cm}$}\psfrag{5.000000e-09}[c][1.0]{$ 0.5\hspace{.3cm}$}\psfrag{1.000000e-08}[c][1.0]{$ 1.0\hspace{.3cm}$}\psfrag{1.500000e-08}[c][1.0]{$ 1.5\hspace{.3cm}$}\psfrag{2.000000e-08}[c][1.0]{$ 2.0\hspace{.3cm}$}\psfrag{2.500000e-08}[c][1.0]{$ 2.5\hspace{.3cm}$}\psfrag{3.000000e-08}[c][1.0]{$ 3.0\hspace{.3cm}$}
\psfrag{Cl}[1.0]{$C_{\ell}(\Delta\nu)\hspace{.1cm}10^{-8}\hspace{.1cm}[\textrm{mK}^2]$}
\psfrag{freq. seper.}[c][1.0]{$\Delta\nu\hspace{.2cm}[\textrm{MHz}]$}
\psfrag{l}[1.0]{$\ell$}
\psfrag{ 0.01}[c][1.0]{$ 0.01$}\psfrag{ 0.1}[c][1.0]{$ 0.1$}\psfrag{ 1}[c][1.0]{$ 1$}\psfrag{ 10}[c][1.0]{$ 10$}
\psfrag{10}{$ $}
\psfrag{2}[1.0]{$10^{2}$}\psfrag{3}[1.0]{$10^{3}$}\psfrag{4}[1.0]{$10^{4}$}
\psfrag{1.000000e+06}[c][c]{$1.0$}\psfrag{1.000000e+05}[c][c]{$0.1$}\psfrag{1.000000e+04}[c][c]{$0.01$}
\includegraphics[scale=0.8,angle=0]{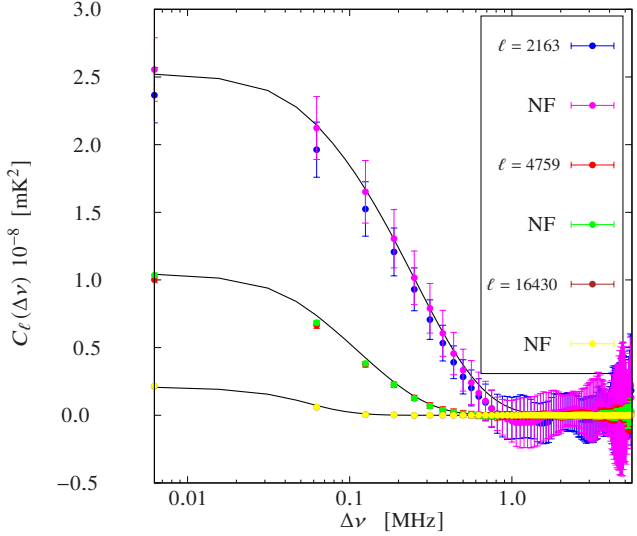}
\caption{This shows $C_{\ell}(\Delta\nu)$ as a function of $\Delta\nu$ for three values of $\ell$. We consider two scenarios: the flagging is identical to the data and no flagging present in the data (denoted by `NF'). The data points with $1\,\sigma$ error-bars are estimated from simulations of 16 different sky realizations drawn from the power spectrum. Note that the $\Delta \nu=0$ points have been slightly shifted for convenience of plotting on a logarithmic scale. The solid black lines show the theoretical predictions calculated by using the input model power spectrum $P^m(k)$.}
\label{cl}
\end{center}
\end{figure}

\begin{figure}
\begin{center}
\psfrag{K2 Mpc3}[1.0]{$\hspace{2cm}P(k_{\perp},k_{\parallel})\hspace{.2cm} \textrm{mK}^{2} \textrm{Mpc}^{3}$}
\psfrag{All channels, avg. 24 realizations, 80\% FLAGGING, taubin}{$ $}
\psfrag{ 0.01}[l][1.0]{$ 0.01$}\psfrag{ 0.1}[l][1.0]{$ 0.1$}\psfrag{ 1}[1.0]{$ 1.0$}
\psfrag{10}{$ $}
\psfrag{-1}[1.0]{$ $}\psfrag{-2}[bc][tl][1.0]{$10^{-2}$}\psfrag{0}[c][l][1.0]{$10^{0}$}\psfrag{1}[1.0]{$ $}\psfrag{2}[c][l][1.0]{$10^{2}$}\psfrag{3}[1.0]{$ $}\psfrag{4}[c][l][1.0]{$10^{4}$}
\psfrag{ 0.4}[c][l][1.0]{$0.4$}\psfrag{-0.4}[c][l][1.0]{$-0.4$}\psfrag{ 0}[r][l][1.0]{$0.0$}\psfrag{ 0.2}[l][l][1.0]{$ $}\psfrag{-0.2}[l][l][1.0]{$ $}
\psfrag{P(k)}[
1.0]{$P(k)\hspace{.2cm} [\textrm{mK}^{2}\textrm{Mpc}^{3}]$}
\psfrag{PM(k)-P(k)/PM(k)}[c][1.0]{$\delta$}
\psfrag{k}[1.0]{$k\hspace{.2cm}[\textrm{Mpc}^{-1}]$}
\psfrag{ 0.01}[1.0]{$ 0.01$}\psfrag{ 1}[1.0]{$ 1.0$}
\psfrag{data}[c][l][0.8]{$\textrm{F} $}
\psfrag{nf}[c][l][0.8]{$\textrm{NF} $}
\psfrag{model}[c][l][0.8]{$\textrm{Model}$}
\includegraphics[scale=.6,angle=0]{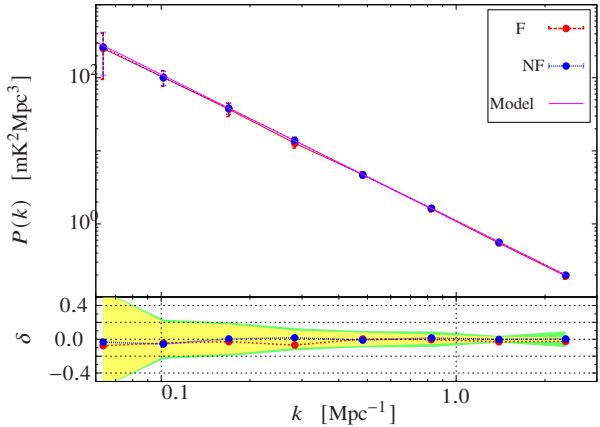}\vspace{3mm}
\caption{The upper panel shows the estimated spherically-binned power spectrum $P(k)$ and $1\,\sigma$ error-bars for the simulation. For `F' the  flagging  is identical to the data  and `NF' has no flagging.
For comparison, the input model $P^{m}(k)$ is also shown by the solid line. The bottom panel shows the fractional error $\delta=[P(k)-P^m(k)]/P^m(k)$  (data points)  and the relative statistical fluctuation $\sigma/P^m(k)$ (shaded regions) with  green and  yellow showing F and NF respectively.}
\label{pk}
\end{center}
\end{figure}

In this section we present  simulations to validate our estimator.
As mentioned earlier, the MAPS based 3D power spectrum estimator, as defined in its previous form in Paper I, has already been  validated using 
$150 \, {\rm MHz}$ GMRT simulations in Paper I.  Here we have repeated the simulations incorporating the particular baseline distribution, flagging and  slightly different central frequency of the  present observation. Note that the present observation has a very sparse baseline distribution corresponding to the  very short observation time, and is rather heavily flagged.  
The aim here is to verify if the new estimator (eq.~\ref{eq:a4}) can still accurately recover the redshifted 21-cm power spectrum in the hypothetical situation where  foregrounds and system noise are absent. Further, the DFT (eq.~\ref{eq:b1}) used in Paper I has now been replaced with the ML (eq.~ \ref{eq:ML}) which is  validated here.

The simulations were carried out on a $[2048]^3$ cubic grid of spatial  resolution  $1.07$ Mpc. This corresponds to an angular  resolution 
of $\sim 23.9^{''}$  and frequency resolution of $\sim 62.5 \, {\rm KHz}$. We assume that the sky signal is described by a $3$D input model brightness temperature power spectrum $P^{m}(k) = (k/k_{0})^n \, {\rm  mK^{2} \,  Mpc^{3}}$, with $k_{0}=(1.1)^{-1/2} \, {\rm  Mpc}^{-1}$ and $n=-2$. We use this to generate multiple random realizations of the sky signal
$\delta T_{\rm b} (\hat{\bm{n}},\,\nu)$ which are used  to simulate the visibilities. In order to validate the estimator, 
the simulated visibilities were analysed in exactly the same way as the actual data. 

Figure \ref{cl} shows the mean estimated $C_{\ell}(\Delta\nu)$ with $1\,\sigma$ error bars
at three values of the angular multipole $\ell$ for the tapering parameter, $f=0.6$. 
We have used $16$ independent  realisations of the simulations to estimate the mean and standard  deviation. Along with this we have also shown the corresponding theoretical model prediction where $C_{\ell}(\Delta\nu)$ is estimated by the taking a Fourier transform of $P^{m}(k)$ (the inverse of eq. (\ref{eq:b1})) along $k_{\parallel}$. We see that the theoretical predictions are within $1\,\sigma$ of the estimated values with the exception of a few points at $\ell=4759$. At $\ell=2163$ and $4759$, the estimator underestimates the $C_{\ell}(\Delta\nu)$ value at small frequency separations. To investigate the source of this discrepancy, we additionally consider a scenario where this same baseline distribution have no flagging. In absence of any flagging, the estimator closely follows the theoretical values at $\ell=2163$ at smaller frequency separations. However, the values are still underestimated at $\ell=4747$. The fractional deviation from the model prediction at $\Delta\nu=0$ is around $-6.15\%,\,-3.85\%\,\textrm{and}\,3.84\%$ at $\ell=2163,\,4759\,\textrm{and}\,16480$ respectively  when flagging is present in the data. For a fixed $\ell$ the deviations at different $\Delta \nu$ appears to be correlated similar to the signal, however the deviations at different $\ell$ appear to be uncorrelated. The exact origin of these small deviations is currently unknown to us. 

We have implemented eq. (\ref{eq:ML}) to estimate the power spectrum of the simulated sky signal. For the simulations we have used the variance of the estimated $C_{\ell}(\Delta\nu)$ as the 
 noise covariance matrix $\textbf{N}$. 
  The pink solid line in the upper panel of Figure \ref{pk} shows the model power spectrum $P^{m}(k)$. The 
  estimated power spectrum with $1\,\sigma$ error bars is shown with 
  blue and red points  showing  the  results without and with flagging respectively. 
   We see that the model power spectrum is within $1\,\sigma$ of the estimated values for the entire  $k$ range.  The lower panel of Figure \ref{pk} show the fractional deviation ($\delta=P(k)-P^{m}(k)/P^{m}(k)$) of the estimated power spectrum for the two cases. The green shaded region shows the $\pm 1\,\sigma$ region   for the flagged data while the yellow shaded region shows the same for the  data without flagging. In both cases, the fractional deviation is within the 
   $\pm 1\,\sigma$ region  for the entire $k$ range. As expected, the $\pm 1\,\sigma$ region 
 is smaller for the unflagged data as compared to its  flagged counterpart. The same is also true for the fractional deviation $\delta$. In case of the flagged data, we see that the fractional deviation is $< 8\%$ over the entire $k$ and lies between $0.2\%$-$-2.6\%$ at $k\,\geq\,0.48$ Mpc$^{-1}$. We conclude that our estimator successfully recovers the input power spectrum, even in presence of the flagging in the data.

\section{Results}
\subsection{The Estimated MAPS}
\label{maps}

\begin{figure*}
\begin{center}
\psfrag{freq. seper.}[c][c][1.5]{$\Delta\nu$ [MHz]}\psfrag{Cl}[r][l][1.5]{$C_{\ell}(\Delta\nu)$ [$\times 10^{3}$ mK$^{2}$]}
\psfrag{522}[c][c][1.2]{$\ell=522$}\psfrag{1065}[c][c][1.2]{$\ell=1065$}\psfrag{3624}[c][c][1.2]{$\ell=3624$}\psfrag{10032}[c][c][1.2]{$\ell=10032$}
\psfrag{f--1}[c][l]{$f=0.4$}\psfrag{f--2}[c][l]{$f=0.6$}\psfrag{f--3}[c][c]{$0.8$}\psfrag{f--4}[c][c]{$2.0$}\psfrag{f--5}[c][c]{$10.0$}
\psfrag{ 0}[c][c][1.5]{$0$}\psfrag{ 1e+06}[c][c][1.5]{ $1$}\psfrag{ 2e+06}[c][c][1.5]{ $2$}\psfrag{ 3e+06}[c][c][1.5]{ $3$}\psfrag{ 4e+06}[c][c][1.5]{ $4$}\psfrag{ 5e+06}[c][c][1.5]{ $5$}\psfrag{ 6e+06}[c][c][1.5]{ $6$}
\psfrag{-0.5}{$ $}\psfrag{ 0.5}{$ $}\psfrag{ 1.5}{$ $}\psfrag{ 2.5}{$ $}\psfrag{ 3.5}{$ $}\psfrag{ 4.5}{$ $}
\psfrag{ 1}[c][c][1.5]{$1$}\psfrag{ 2}[c][c][1.5]{$2$}\psfrag{ 3}[c][c][1.5]{$3$}\psfrag{ 4}[c][c][1.5]{$4$}
\psfrag{-2}[c][c][1.5]{$-2$}\psfrag{-4}[c][c][1.5]{$-4$}\psfrag{-5}[c][c][1.5]{$ $}\psfrag{-1}[c][c][1.5]{$ $}\psfrag{ 7}[c][c][1.5]{$ $}\psfrag{-10}[c][c][1.5]{$-10$}\psfrag{ 30}[c][c][1.5]{$30$}\psfrag{-15}[c][c][1.5]{$ $}\psfrag{ 35}[c][c][1.5]{$ $}
\psfrag{ 6}[c][c][1.5]{$6$}\psfrag{ 8}[c][c][1.5]{$8$}\psfrag{ 10}[c][c][1.5]{$10$}\psfrag{ 12}[c][c][1.5]{$12$}
\psfrag{ 14}[c][c][1.5]{$14$}\psfrag{ 16}[c][c][1.5]{$16$}\psfrag{ 18}[c][c][1.5]{$18$}\psfrag{ 20}[c][c][1.5]{$20$}
\psfrag{ 5}[c][c][1.5]{$ $}\psfrag{ 15}[c][c][1.5]{$ $}\psfrag{ 25}[c][c][1.5]{$ $}
\includegraphics[width=140mm]{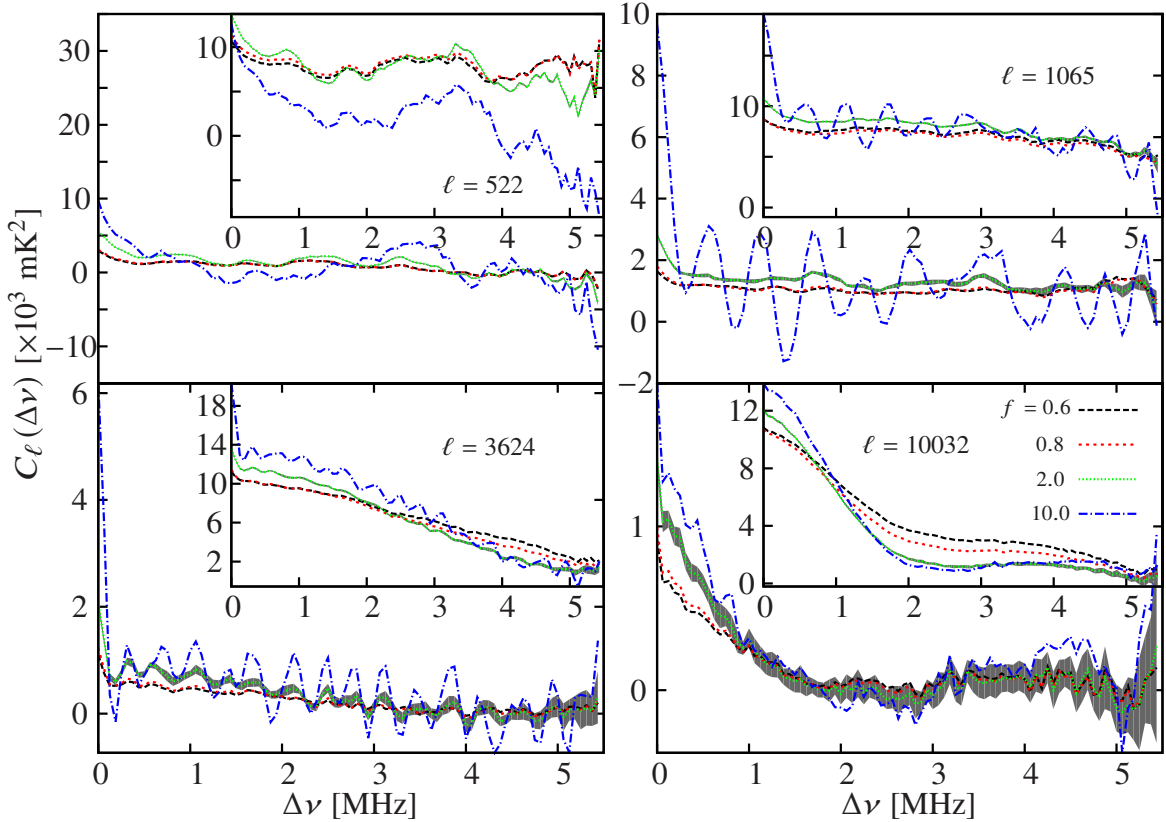}\vspace{10mm}
\caption{$C_{\ell}(\Delta\nu)$ as a function of $\Delta\nu$ after point source subtraction, with  before point  source subtraction  shown as inset. The different panels correspond to different values of $\ell$, and the different lines correspond to different $f$ values as indicated in the legend. 
The black shaded regions for $f=2.0$ displays the $10\,\sigma$ error bars due to the system noise only.}
\label{Cl_taper}
\end{center}
\end{figure*}

We have estimated the MAPS $C_{\ell}(\Delta\nu)$ directly from the visibilities using eq. (\ref{eq:a4})  using  the calibrated visibilities both  before and after  point source subtraction. As mentioned earlier, we have used $88$ frequency channels from the central region of the frequency band, two polarizations (RR and LL) and a baseline range of $ 70 \lambda \leq \lvert {\bf U}_{i} \rvert \leq 3000 \lambda$ for the analysis. The Gaussian window function (eq.~\ref{eq:win1}) is adopted to taper the GMRT PB pattern and we have considered four values of the tapering parameter `$f$' for this analysis namely  $f=0.6,\,0.8,\,2.0\, \textrm{and}\,10.0$. Tapering increases with decreasing value of `$f$', and $f=10.0$ is equivalent 
to an untapered PB pattern. We have generated $20$ realizations of the UMAPS and the corresponding simulated visibilities
were used to estimate the normalization factor $M_{g}$. We have binned   $C_{\ell}(\Delta\nu)$ into $15$ logarithmic bins along $\ell$. We note that the convolution in eq. (\ref{eq:a1}) is expected to be important at large angular scales (small $\ell$), and the extent of this $\ell$ range increases as $f$ is decreased. \citet{samir14} have studied this in detail using simulations. Their results indicate that the effect of the convolution is restricted to small $\ell$, and  we may ignore the effect of the convolution at  large multipoles $\ell \ge \ell_{min}$  where $\ell_{min} \approx 13.3\,  \sqrt{1+f^2} \,/ (f*\theta_{FWHM})$. We have used this  to  account for the convolution  by discarding the multipoles $\ell < \ell_{min}$ when binning  the estimated power spectrum. As a consequence the
smallest $\ell$ values which are accessible are approximately  $\ell_{min}= 563,\,463,\,324,\,291$ for f= $0.6,\,0.8,\,2.0,\,10.0$ respectively. The 
$\ell$ value corresponding to each $\ell$ bin also changes to some extent with the tapering  $f$. We note that our  critereon  based on $\ell_{min}$ is  rather approximate in that the exact extent and effect of the convolution is sensitive to the $\ell$  dependence (slope)  of $C_{\ell}(\Delta \nu)$. It is possible that,  for the particular signal in our simulations or in the observational data,  the effect of the convolution  extends beyond $\ell_{min}$ into a few of the smallest $\ell$ bins which we have used for our analysis. A more precise power spectrum estimation would involve  deconvolving the effects of the primary beam pattern and the tapering window, however the present approach is adequate given the high   
noise level of the present data.

Figure \ref{Cl_taper} shows the binned $C_{\ell}(\Delta\nu)$ over a bandwidth of $5.5$ MHz before (panels shown in the inset) and after the point sources are subtracted. We show the results for the  tapering values $f=0.6,\,0.8,\,2.0\,\textrm{and}\,10.0$. The black shaded regions for $f=2.0$ displays $10 \times 
[\delta C_{\ell}(\Delta\nu)]$  where $ [\delta C_{\ell}(\Delta\nu)]$ refers to  the 
estimated statistical errors due to the system noise only. The measured visibilities are system noise dominated, and we have 
 used the real and imaginary parts of the measured visibilities to  estimate the variance $\sigma^{2}_{N}$.
 The measured $\sigma^{2}_{N}$ values are $(3.87)^{2}\, \textrm{Jy}^{2}$ and $(3.42)^{2}\, \textrm{Jy}^{2}$ before and after point source subtraction  respectively. We have  simulated $20$ realisations of the  visibilities corresponding to  Gaussian random fluctuations with zero mean and variance $\sigma^{2}_{N}$. The MAPS estimator we applied to the simulated visibilities, and the $C_{\ell}(\Delta\nu)$ estimated from the $20$ realisations were  used to determine the variance $ [\delta C_{\ell}(\Delta\nu)]^2$. We have also estimated the power spectrum  from the noise simulations and used these to estimate the variance $[\delta  P_{N}]^2$ arising from the system noise.

We notice that the estimated $C_{\ell}(\Delta\nu)$ remain correlated over
the analysed bandwidth ($5.5$ MHz) at small $\ell$'s and decorrelates relatively faster at the larger  $\ell$ bins. 
Considering  any  fixed $\ell$ bin, the decorrelation with $\Delta\nu$ is faster  after the point sources have been subtracted. 
The overall amplitude of $C_{\ell}(\Delta\nu)$ falls approximately by one order of magnitude, especially at higher $\ell$ values when the point sources are removed. An oscillatory pattern is also observed at all angular scales for both sets of data. These observed oscillations are due to the strong point sources located away from the phase center of the observations, 
consistent with the previous results  reported in \citet{ghosh3}. The frequency of the oscillations is found to increase at larger baselines (higher $\ell$ values). The tapering of the PB pattern suppresses the contributions from  the outer parts of the FoV which  brings down the amplitude of  the oscillations. In order to quantify the effect of tapering we first focus on the estimated $C_{\ell}(\Delta\nu_c)$ {\it i.e.} a single channel separation at $\ell=1065$.
 We find  that relative to $f=10.0$,  the amplitude of $C_{\ell}(\Delta\nu)$ falls by a factor of $1.66,\,2.03,\,\textrm{and}\,2.02$ for $f=2.0,\,0.8,\,\textrm{and}\,0.6$ respectively 
before point source subtraction. The corresponding values are $3.27,\,5.01,\,\textrm{and}\,5.40$  after point source subtraction. Considering $C_{\ell}(\Delta\nu_c)$, this factor is overall  $< 1.7, 2.1, 2.2$ at $f=2.0,\,0.8,\,\textrm{and}\,0.6$ respectively for all $\ell$ before point source subtraction, and it is  $< 5.5, 9.8, 10.2$ at $f=2.0,\,0.8,\,\textrm{and}\,0.6$ respectively after point source subtraction. The suppression is also  found to somewhat saturate beyond  $f=0.8$, and there is not much  improvement if $f$ is reduced further. 
 The suppression due to tapering is also visible in the estimated values of $P(k_{\perp},k_{\parallel})$ which we shall discuss later in the Section \ref{powerspec}.

\begin{figure}
\begin{center}
\psfrag{freq. seper.}[c][c][1.0]{$\Delta\nu$ [MHz]}\psfrag{Cl}[c][c][1.0]{$C_{\ell}(\Delta\nu)$ [$\times 10^{3}$ mK$^{2}$]}
\psfrag{852}[cb][ct][1.0]{$\ell=852$}\psfrag{10032}[c][c][1.0]{$\ell=10032$}
\psfrag{f--1}[c][l]{$f=0.4$}\psfrag{f--2}[r][c]{$f=0.6$}\psfrag{f--3}[c][l]{$0.8$}\psfrag{f--4}[c][c]{$2.0$}\psfrag{f--5}[c][c]{$10.0$}
\psfrag{ 0}[c][c][1.0]{$0$}\psfrag{ 1e+06}[c][c][1.0]{ $1$}\psfrag{ 2e+06}[c][c][1.0]{ $2$}\psfrag{ 3e+06}[c][c][1.0]{ $3$}\psfrag{ 4e+06}[c][c][1.0]{ $4$}\psfrag{ 5e+06}[c][c][1.0]{ $5$}\psfrag{ 6e+06}[c][c][1.0]{ $6$}
\psfrag{-0.5}{$ $}\psfrag{ 0.5}{$ $}\psfrag{ 1.5}{$ $}\psfrag{ 2.5}{$ $}\psfrag{ 3.5}{$ $}\psfrag{ 4.5}{$ $}
\psfrag{ 1}[c][c][1.0]{$1$}\psfrag{ 2}[c][c][1.0]{$2$}\psfrag{ 3}[c][c][1.0]{$3$}\psfrag{ 4}[c][c][1.0]{$4$}
\psfrag{-2}[c][c][1.5]{$ $}
\psfrag{ 6}[c][c][1.0]{$6$}\psfrag{ 8}[c][c][1.0]{$8$}\psfrag{ 10}[c][c][1.0]{$10$}\psfrag{ 12}[c][c][1.0]{$12$}
\psfrag{ 14}[c][c][1.0]{$14$}\psfrag{ 16}[c][c][1.0]{$16$}\psfrag{ 18}[c][c][1.0]{$18$}\psfrag{ 20}[c][c][1.0]{$20$}
\psfrag{ 5}[c][c][1.0]{$ $}\psfrag{ 15}[c][c][1.0]{$ $}\psfrag{ 25}[c][c][1.0]{$ $}
\includegraphics[width=70mm]{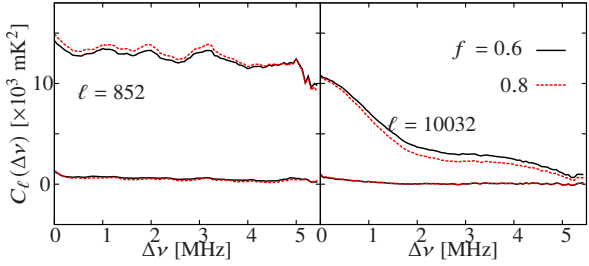}
\caption{This shows $C_{\ell}(\Delta\nu)$ as a function $\Delta\nu$ before and after point source subtraction for two $\ell$ values (different  panels) and  tapering parameters $f=0.6 \,\textrm{and}\, 0.8$. }
\label{Cl_tot+res}
\end{center}
\end{figure}

 At a given $\ell$ the oscillations  are more pronounced before the point sources have been subtracted.
 This feature is not quite obvious from Figure \ref{Cl_taper}. Figure \ref{Cl_tot+res} shows the estimated $C_{\ell}(\Delta\nu)$ before and after the point source subtraction for two representative $\ell$ values with  tapering parameter $f=0.8$ and $0.6$. We find that  although the oscillation do not completely go away, they   become much smoother after the point sources have been subtracted. We see  that  at small $\ell$'s the nature of the oscillatory patterns in the residual data is similar to that  before point source subtraction, only the amplitude of oscillations are somewhat smaller for the residual $C_{\ell}(\Delta\nu)$.
The degree of suppression due to tapering depends on the effectiveness of the convolution which, in turn, depends on the baseline distribution. The tapering suppression is expected to be more effective in a situation with more uniform and denser baseline distribution.

\begin{figure*}
\begin{center}
\psfrag{freq. seper.}[c][c][1.5]{$\Delta\nu$ [MHz]}\psfrag{Cl}[r][l][1.5]{$C_{\ell}(\Delta\nu)$ [$\times 10^{3}$ mK$^{2}$]}
\psfrag{522}[l][r][1.2]{$\ell=522$}\psfrag{642}[c][c][1.2]{$\ell=642$}\psfrag{1064}[lc][rb][1.2]{$\ell=1064$}\psfrag{1805}[l][r][1.2]{$\ell=1805$}\psfrag{3624}[l][r][1.2]{$\ell=3624$}\psfrag{4751}[l][r][1.2]{$\ell=4751$}\psfrag{7891}[l][r][1.2]{$\ell=7891$}\psfrag{10032}[l][r][1.2]{$\ell=10032$}
\psfrag{f--2}[r][c]{$f=0.6$}\psfrag{f--3}[c][l]{$0.8$}
\psfrag{ 0}[c][c]{$0$}\psfrag{ 1e+06}[c][c]{ $1$}\psfrag{ 2e+06}[c][c]{ $2$}\psfrag{ 3e+06}[c][c]{ $3$}\psfrag{ 4e+06}[c][c]{ $4$}\psfrag{ 5e+06}[c][c]{ $5$}
\psfrag{ 0.6}{$ $}\psfrag{ 0.8}{$0.8$}\psfrag{ 1}{$ $}\psfrag{ 1.2}{$ $}\psfrag{ 1.4}{$ $}\psfrag{ 1.6}{$1.6$}\psfrag{ 1.8}{$ $}\psfrag{ 2}{$ $}\psfrag{ 2.2}{$ $}
\psfrag{-1.5}[c][l]{$-1.5$}\psfrag{-1}[c][l]{$ $}\psfrag{-0.5}{$ $}\psfrag{ 0.5}{$ $}\psfrag{ 1.5}{$1.5$}\psfrag{ 2.5}{$ $}
\psfrag{-0.2}[c][c]{$ $}\psfrag{-0.4}[c][c]{$ $}\psfrag{-0.6}[c][c]{$ $}
\psfrag{ 0.2}{$ $}\psfrag{ 0.4}{$ $}
\psfrag{-2}[c][l]{$ $}\psfrag{-4}[c][l]{$ $}\psfrag{ 4}[c][l]{$ $}\psfrag{-3}[c][l]{$-3$}\psfrag{ 3}[c][l]{$3$}
\includegraphics[width=140mm]{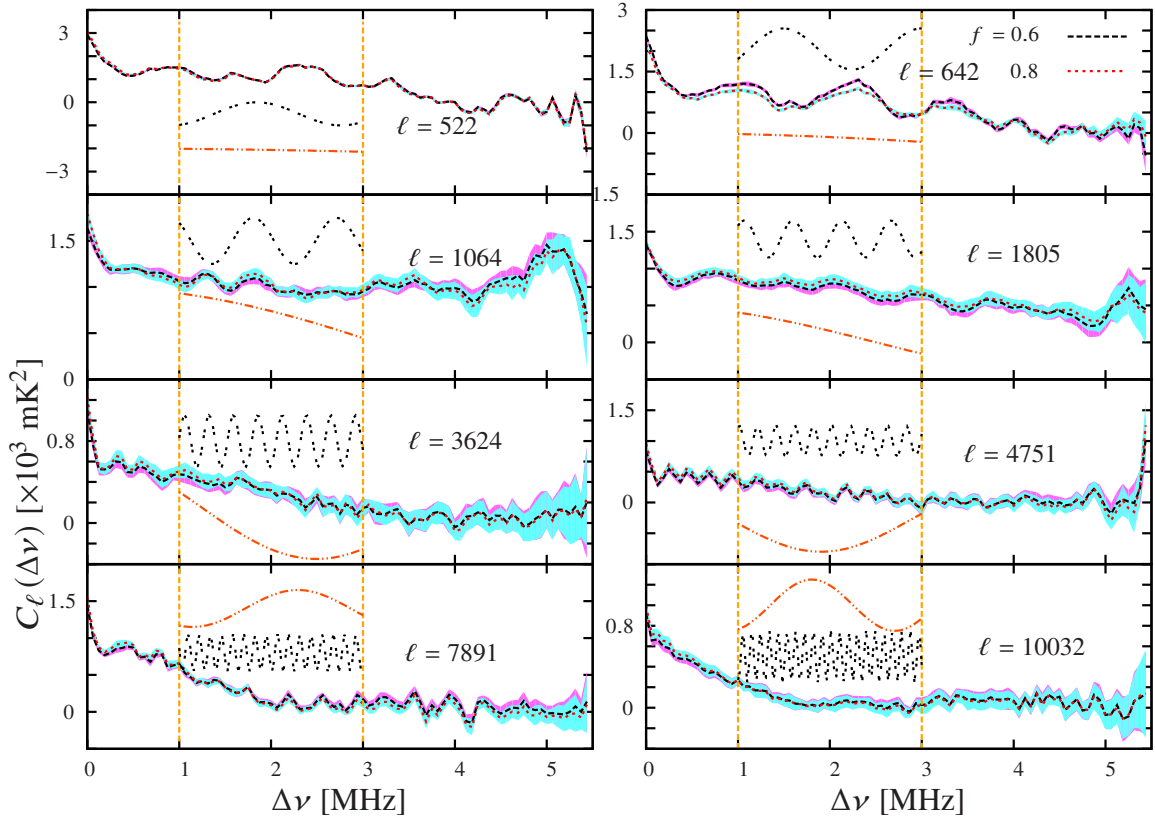}\vspace{10mm}
\caption{ $C_{\ell}(\Delta\nu)$ as a function of $\Delta\nu$ at eight  values of $\ell$ (different panels)
and  two values of $f$ after the point source subtraction. The shaded regions denote the $10\,\sigma$ error due to the system noise only. In each panel, the region bounded by orange dashed vertical lines shows the components, $[C_{\ell}(\Delta\nu)]_{\vtha}$ (orange dashed-dot-dot lines) and $ [C_{\ell}(\Delta\nu)]_{H}$ (black small dashed lines) corresponding to a source located at first null of the PB and the horizon limit respectively as discussed in eqs. (\ref{eq:null}) and (\ref{eq:hori}).  The amplitude of the oscillating components bounded by the orange vertical dashed lines have been adjusted for the convenience of plotting.}
\label{Cl_res}
\end{center}
\end{figure*}

The estimated $C_{\ell}(\Delta\nu)$  is foreground dominated. The analysis till now indicates that the values of the 
 estimated $C_{\ell}(\Delta\nu)$, and the oscillations therein, are both considerably reduced  if we use the data after point source subtraction. Further, the amplitude is also found to decrease if the tapering parameter $f$ is reduced. However, this effect saturates beyond $f=0.8$.  Reducing $f$ also enhances the cosmic variance.  Based on  these considerations we have primarily focused on $f=0.6$  and $0.8$ for the subsequent analysis. Figure \ref{Cl_res} shows the estimated $C_{\ell}(\Delta\nu)$ for these two values of the tapering parameter ($f=0.6$ and $0.8$) after point source subtraction. 
The overall amplitude varies approximately between $0.9\,\times\,10^{3} \, {\rm mK}^{2}$ and  $3.0\,\times\,10^{3} \, {\rm mK}^{2}$ for all the  $\ell$ bins shown in the figure. It is expected that around $\ell \leq 800$ the visibilities (and the derived $C_{\ell}(\Delta\nu)$) are dominated by DGSE, whereas for  $\ell \geq 800$ the residual point sources are found to dominate \citep{ghosh3} .
Various models (e.g. \cite{KD07,Mondal20}) predict the strength of the $21$-cm signal at these $\ell$ values to be around $10^{-5} - 10^{-6} \, {\rm mK}^{2}$ - orders of magnitude lower than the estimated $C_{\ell}(\Delta\nu)$ values. The 
$C_{\ell}(\Delta\nu)$ for the $21$-cm signal, however, is predicted to decorrelate very rapidly as the frequency separation $\Delta \nu$ in increased. It is expected to approximately fall by $50\%$ within $\Delta \nu=0.5$ MHz at $\ell=10^{2}$ and within $\Delta \nu =10$ kHz at $\ell=10^{5}$ respectively. The spectrally smooth foreground contributions (which are largely continuum sources)
 are expected to remain correlated over large $\Delta \nu$.   This difference
is  expected to play a very crucial role in extracting the 21-cm signal from the foregrounds. 
We however notice   oscillatory feature present for all the $\ell$ values shown in Figure \ref{Cl_res}.
These oscillations, whose amplitudes are several  orders of magnitude larger than the 21-cm signal, pose a serious challenge for separating the 21-cm  signal from the foregrounds. 

Considering a point source of flux density $S_{\nu}$ located along unit vector $\mathbf{\hat{n}}$, 
 its contribution to the measured visibility is 
\begin{equation}
    \V_i(\nu)=S_{\nu} \, A(\Delta  \mathbf{\hat{n}},\nu) \, \exp\{ 2 \pi \u_i \cdot  \Delta  \mathbf{\hat{n}}\,
    (\nu/\nu_c) \} \,.
    \label{eq:vis1}
\end{equation}
where $\Delta  \mathbf{\hat{n}}=  \mathbf{\hat{n}}-  \mathbf{\hat{m}}$
 with  $\mathbf{m}$ denoting the unit vector to the phase center. 
We have a net oscillation $\propto  \cos[ 2 \pi \u_i \cdot \Delta \mathbf{\hat{n}}  \, (\nu_a -\nu_b)/\nu_c] $ when we 
correlate visibilities at the same baseline and two different frequencies $\nu_a$ and $\nu_b$ in order to estimate 
$C_{\ell}(\Delta\nu)$ at the frequency separation $\Delta \nu=\mid \nu_a-\nu_b \mid$. These oscillations, whose frequency increases with $\u_i \cdot  \Delta  \mathbf{\hat{n}}$
 are primarily what we see in the measured $C_{\ell}(\Delta\nu)$. These features are mainly caused  by the strong point sources located away from the center of the FoV. Here we explicitly discuss two cases, the first being a source located at $\vtha=\mid \Delta  \mathbf{\hat{n}} \mid  \ll 1$  the first null of the PB in which  case we have have oscillations \begin{equation}
     [C_{\ell}(\Delta\nu)]_{\vtha} \propto \cos ( \ell \vtha \, \Delta \nu/\nu_c) \,. 
\label{eq:null}
 \end{equation}
 Note that we have used $\vtha=3.046^{\circ}$ here. 
 Another case that we consider is a source located at the horizon for which $\u_i \cdot \Delta  \mathbf{\hat{n}}=1$ and we have 
\begin{equation}
     [C_{\ell}(\Delta\nu)]_{H} \propto \cos ( \ell  \, \Delta \nu/\nu_c) \,. 
\label{eq:hori}
 \end{equation}
In addition to the estimated $C_{\ell}(\Delta \nu)$,   
Figure \ref{Cl_res} also shows the oscillations predicted by eq. (\ref{eq:null}) and (\ref{eq:hori}) over the $\Delta \nu$  range $1-3$ MHz  demarcated by the orange dashed vertical lines.  As expected, the oscillation period $[\Delta \nu]_P$ is much larger for $[C_{\ell}(\Delta\nu)]_{\vtha}$ as compared to  $[C_{\ell}(\Delta\nu)]_{H}$. For 
$[C_{\ell}(\Delta\nu)]_{\vtha}$, in most cases  $[\Delta \nu]_P$ is larger than the $\Delta \nu$ interval 
of $1-3 $ MHz and only a fraction of the sinusoidal oscillation is visible in the figure. In all cases, $[\Delta \nu]_P$
decreases and the oscillations get more rapid as $\ell$ is increased. Considering the estimated $C_{\ell}(\Delta \nu)$, we see that the oscillatory patterns are more complex than the simple sinusoidal oscillations in  $[C_{\ell}(\Delta\nu)]_{\vtha}$ and  $[C_{\ell}(\Delta\nu)]_{H}$.  Considering $[\Delta \nu]_P$ which denotes the period of the most dominant component of the oscillations seen in the estimated $C_{\ell}(\Delta\nu)$,  we see that  $[\Delta \nu]_P$ decreases with increasing $\ell$. In most cases  $[\Delta \nu]_P$ for the estimated $C_{\ell}(\Delta \nu)$ is between those of $[C_{\ell}(\Delta\nu)]_{\vtha}$ and  $[C_{\ell}(\Delta\nu)]_{H}$. This indicates that the sources responsible for the oscillations are mainly located between the first null and the horizon. 
We note that the dominant contribution from point source within the FWHM of the PB has been modelled and removed, 
 however the contribution from  point sources at larger angular distances remains. We see that this is manifested in the period $[\Delta \nu]_P$  of the oscillations observed in the estimated 
$C_{\ell}(\Delta \nu)$.  The observed oscillations are a superposition  of the  oscillatory contributions from all the strong point sources outside the FoV of the telescope. Considering  sources above a flux cut-off of 1 Jy, TGSS-ADR1 \citep{Intema17} source catalogue lists $5,\,31\,\textrm{and}\,69$ sources close to the first, second and third null of the GMRT PB respectively for the present FoV.

In addition to this, the  PB pattern $A(\mathbf{\theta},\nu)$ changes with frequency, 
 and the angular positions of the nulls and the side-lobes change with frequency. Bright continuum sources located near the nulls or in the sidelobes will be perceived as oscillations
along the frequency axis in the measured visibilities. It is
thus quite likely that these bright sources produce additional oscillatory features  in
the measured $C_{\ell}(\Delta\nu)$.
 Recent LOFAR 21-cm signal upper limits \citep{mertens20} have also found an excess power due to spectral structure with a coherence scale of 0.25 - 0.45 MHz.  This could be due to residual foreground emission from sources or diffuse emission far away from the phase centre, polarization leakage, or low-level radio-frequency interference. Tapering  the array's sky response suppresses the sidelobe response, and  it is possible  that these problem can be mitigated \citep{ghosh2} by adopting such an approach.
 
\subsection{3D Power Spectrum}
\label{powerspec}
\begin{figure}
\psfrag{kpara in Mpc-1}[c][c][1.5]{$k_{\parallel}$ [Mpc$^{-1}$]}\psfrag{kper in Mpc-1}[t][b][1.5]{$k_{\perp}$ [Mpc$^{-1}$]}\psfrag{K2 Mpc3}[c][t][1.2]{[K$^{2}$ Mpc$^{3}$]}
\psfrag{f2}[r][r][1.5]{$f=0.6$}\psfrag{f3}[r][r][1.5]{$0.8$}\psfrag{f4}[r][r][1.5]{$2.0$}\psfrag{f5}[r][r][1.5]{$10.0$}
\psfrag{k1}[t][b][1.5]{$k_{\perp}\approx0.08$}\psfrag{k2}[lt][rb][1.5]{$k_{\perp}\approx0.17$}\psfrag{k3}[lt][rb][1.5]{$k_{\perp}\approx0.5$}
\psfrag{ 0.01}[c][c][1.5]{$ $}\psfrag{ 0.1}[c][c][1.5]{$0.1$}\psfrag{ 1}[c][c][1.5]{$1$}
\psfrag{10}{$ $}
\psfrag{2}[r][cb][1.5]{$10^{2}$}\psfrag{3}[r][c][1.5]{$10^{3}$}\psfrag{4}[r][c][1.5]{$10^{4}$}\psfrag{5}[r][c][1.5]{$10^{5}$}\psfrag{6}[r][c][1.5]{$10^{6}$}\psfrag{7}[r][c][1.5]{$10^{7}$}\psfrag{8}[r][c][1.5]{$10^{8}$}
\psfrag{ 2}[r][c][1.5]{$2$}\psfrag{ 4}[r][c][1.5]{$ $}\psfrag{ 6}[r][c][1.5]{$6$}\psfrag{ 8}[r][c][1.5]{$ $}\psfrag{ 10}[r][c][1.5]{$10$}\psfrag{ 12}[r][c][1.5]{$ $}\psfrag{ 14}[r][c][1.5]{$14$}
\psfrag{ratio}{$ $}
\psfrag{U}[cb][ct][1.0]{$\ell$}\psfrag{147}[cb][ct][1.5]{$147$}\psfrag{1470}[cb][lt][1.5]{$1470$}\psfrag{0.027}[c][c][1.5]{$ $}\psfrag{0.27}[c][c][1.5]{$0.27$}\psfrag{2.7}[c][c][1.5]{$2.7$}\psfrag{Tau}[cb][rt][1.0][270]{$\tau$ [ $\mu$s ]}
\includegraphics[width=90mm]{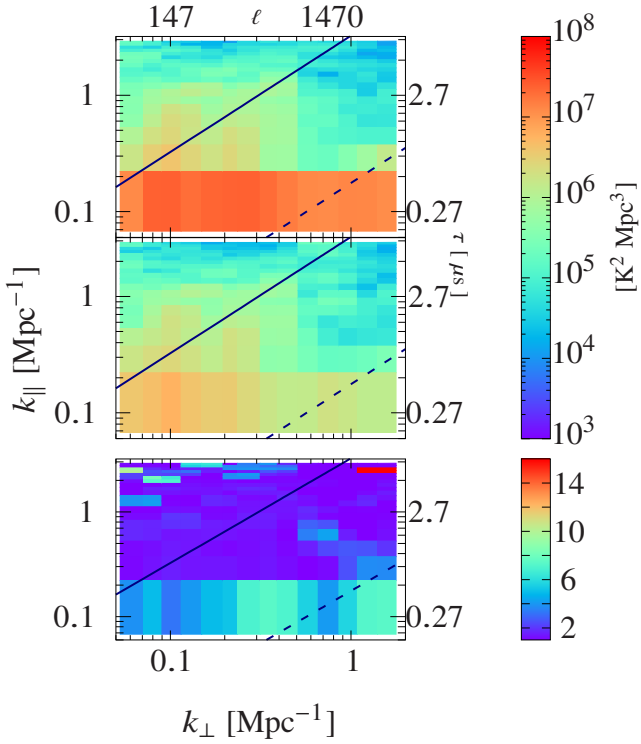}\vspace{6mm}
\caption{The absolute values of the estimated cylindrically-binned power spectrum $P(k_{\perp},k_{\parallel})$ are shown before (uppermost panel) and after (middle panel) the point source subtraction, for tapering parameter $f=10.0$. The lowermost panel shows the ratio of the two power spectra before and after the point sources have been removed. In all panels the blue  dashed  and solid lines  respectively correspond to the first null and the horizon limit of the foreground wedge.}
\label{pk_10}
\end{figure}

We have used the maximum likelihood technique (eq. \ref{eq:ML}) to 
estimate the 3D power spectrum $P(k_{\perp},\,k_{\parallel})$ from the 
$C_{\ell}(\Delta\nu)$ presented  in Section \ref{maps}.  We have used $ [\delta C_{\ell}(\Delta\nu)]^2$ 
estimated from the noise simulations for the  noise covariance matrix $\textbf{N}$ which is expected to be diagonal.
The upper and middle panels of Figure \ref{pk_10} respectively present the absolute value of the binned cylindrical  power spectra ($\mid P(k_{\perp},\,k_{\parallel}) \mid $)
before and after point source  subtraction with  $f=10.0$ which essentially corresponds to no tapering. 
Each $k_{\perp}$ bin here corresponds to an $\ell$ bin of $C_{\ell}(\Delta\nu)$ 
with  $k_{\perp}=\ell/r$.  The $k_{\parallel}$ range has been divided in twenty linear bins of equal width. 
Any feature in $C_{\ell}(\Delta\nu)$ with a period $[\Delta \nu]_P$ is reflected as a feature in $P(k_{\perp},\,k_{\parallel})$ at $k_{\parallel}=2 \pi/([\Delta \nu]_P \, r^{\prime})$.  Consider $[C_{\ell}(\Delta\nu)]_{\vtha}$ and $[C_{\ell}(\Delta\nu)]_{H}$
( eq.~\ref{eq:null} and eq.~\ref{eq:hori}) which respectively have  periods $[\Delta \nu_P]_{\vtha}=2 \pi \nu_c/(\ell \, \vtha))$ and  $[\Delta \nu_P]_{\vtha}=2 \pi \nu_c/\ell$. 
These  $\Delta \nu_P$ which vary with $\ell$ correspond to the straight lines $[k_{\parallel}]_{\vtha} = (r\vtha/r^{\prime}\nu_{c}) \, k_{\perp}$ and $[k_{\parallel}]_{H} = (r/r^{\prime}\nu_{c}) \, k_{\perp}$  which are  also respectively shown in  Figure \ref{pk_10}. 
The foreground contributions from sources located within the first null will appear within  $k_{\parallel} \le [k_{\parallel}]_{\vtha}$ provided we ignore the intrinsic spectral variations of the foreground sources and the chromatic response   of the PB. Under the same conditions we expect the entire foreground contribution to be restricted within $k_{\parallel} \le [k_{\parallel}]_H$, the so called `Foreground Wedge', creating the `EoR  Window'  for redshifted 21-cm HI studies at higher $k_{\parallel}$ values outside the wedge.   However, in reality the foreground sources and PB both exhibit spectral structures which lead to foreground leakage  outside the wedge. Typically one needs to also avoid a 
$k_{\parallel}$ region above  the wedge boundary due to the the leakage.  

Considering Figure \ref{pk_10} we see that both before and after point source subtraction  the foreground contributions are largely confined within the foreground wedge,  however there is also some foreground leakage to $k_{\parallel}$  modes beyond the wedge. In an earlier study \citet{ghosh3} have shown that  point sources  are the dominant foreground component at all the angular multipoles here before point source subtraction, whereas after point source subtraction the DGSE dominates at $\ell < 800$ (the lowest $ k_{\perp}$ bin here) while  the point sources continue to dominate at larger $\ell$. 
We find (Figure \ref{pk_10}) that the leakage outside the foreground wedge is  most prominent in the range 
$0.09 \, \rm{Mpc^{-1}} < k_{\perp} < 0.5 \, \rm{Mpc^{-1}}$  which  is   point source dominated . Considering the upper panel we find that the foreground power is particularly large 
($\sim 2-4 \times 10^7  \, \rm{K^2  \, Mpc^3}$) within
$k_{\parallel} \le 0.2 \, \rm{Mpc^{-1}}$ across all $k_{\perp}$. At large $k_{\parallel}$ outside the wedge 
the power  fall by $2-4$ orders of magnitude (to values in the range $\sim 10^{3}\, {\rm K^2 \, Mpc^3}$ (at larger $k_{\perp}$)  $- 10^{5} \, {\rm K^2 \, Mpc^3}$ (at $k_{\perp} < 0.09 \, {\rm Mpc}^{-1}$)) where it becomes comparable to the noise. We also notice a region with $k_{\perp} \ge 0.5 \, {\rm Mpc}^{-1}$ where the power falls to values in the range $\sim10^{3}-10^{5} \, {\rm K^2 \, Mpc^3}$ at  $k_{\parallel}>0.2 \, {\rm Mpc}^{-1}$ even inside the wedge. The overall structure remains the same after point source subtraction (middle panel).  We notice a  drop in power to values 
$\sim (1-10\times10^{6})  \, {\rm K^2 \, Mpc^3}$ at $k_{\parallel} \le 0.2 \, \rm{Mpc^{-1}}$, the amplitude of the leakage power is also found to be lower compared to before point source subtraction.  
The lowermost panel of Figure \ref{pk_10} shows the ratio of power before and after source subtraction. We find 
that this ratio has values $< 15$  at $k_{\parallel} \le 0.2 \, \rm{Mpc^{-1}}$ across the entire  $k_{\perp}$ range.
The ratio is of order unity elsewhere, including the  EoR window and the first $k_{\perp}$ bin which is expected to be DGSE dominated after point source subtraction, with the exception of a very few $k_{\parallel}$ bins at the higher end.

\begin{figure*}
\psfrag{1}[c][c][1.4]{$1$}\psfrag{10}{$ $}\psfrag{4}[rc][lb][1.4]{$10^{4}$}\psfrag{6}[rc][lb][1.4]{$10^{6}$}
\includegraphics[scale=1.1]{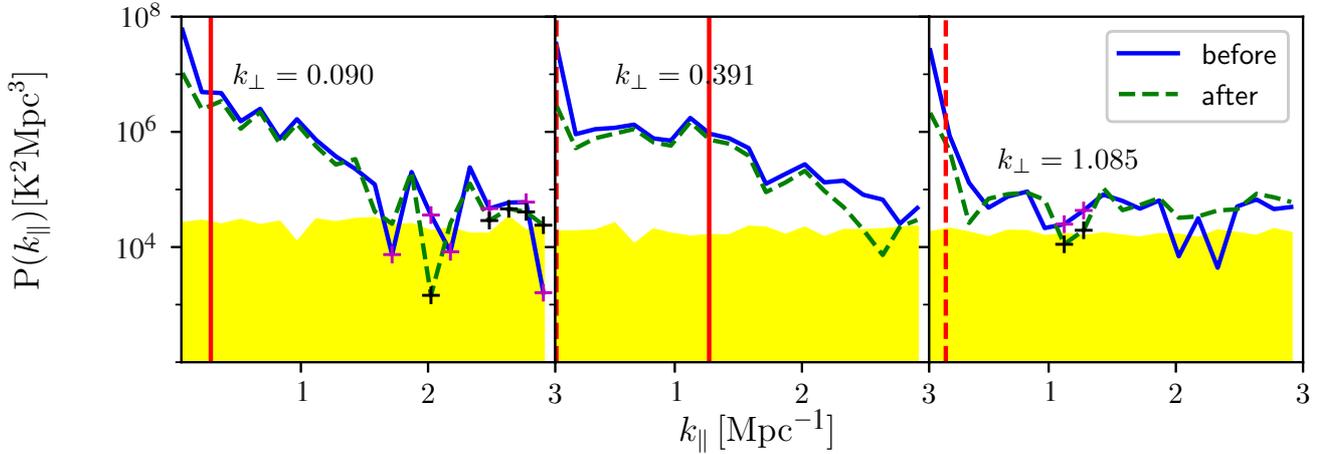}
\caption{The absolute value of the estimated cylindrically-binned power spectrum $P(k_{\perp},k_{\parallel})$ before (blue solid)  and after (green dashed) point source subtraction for tapering parameter $f=10.0$, as a function of $k_{\parallel}$, at three values of $k_{\perp}$. The $k_{\parallel}$ bins corresponding to the negative power values are indicated by magenta and black `+' markers before and after the point source subtraction respectively. The shaded regions show the $1\,\sigma$ errors $\delta P_{N}$ due to the system noise after point source subtraction. In all the cases, the solid  and dashed red vertical lines respectively denote  $[k_{\parallel}]_{H}$
and $[k_{\parallel}]_{\vtha}$.}
\label{pk_10_slice}
\end{figure*}

Figure \ref{pk_10_slice} shows  $\mid P(k_{\perp},\,k_{\parallel}) \mid $ as a function of  $k_{\parallel}$ for three representative values of $k_{\perp}$. The shaded regions shows  the $1\,\sigma$ errors  $(\delta  P_{N})$ due to the system noise  after point source subtraction. We expect the extent of the shaded region to increase by a factor of $\sim 1.3$ before point source subtraction. This factor corresponds to the ratio of the $\sigma^2_N$ values before and after the point source subtraction. We find that the power is maximum at the the lowest $k_{\parallel}$, and it has values in the range $2.6-6 \times 10^{7}\,\textrm{K}^{2}\textrm{Mpc}^{3}$ before point source subtraction. The power falls at higher $k_{\parallel}$ and  the power drops by a factor of $\sim 10^{4}$ at  $k_{\perp}=0.090 \, {\rm Mpc}^{-1}$.  The roll-off  is steeper at larger $k_{\perp}$, and the power  drops by a factor  $\sim 10^3$ at $k_{\perp}=1.085 \, {\rm Mpc}^{-1}$ within $k_{\parallel}\sim 0.5\,{\rm Mpc}^{-1}$.
For most $k_{\perp}$  the  power  becomes comparable to the noise and exhibits both positive and negative values at  $k_{\parallel}\geq 1.5 \,\textrm{Mpc}^{-1}$. The $k_{\parallel}$ values corresponding to the negative power are indicated by `+' in Figure \ref{pk_10_slice} before and after the point source subtraction. The behaviour at large $k_{\parallel}$ does not change much if point source are subtracted. However, the power falls by a factor of $\sim 10$ at the lowest $k_{\parallel}$. Typically the difference between before and after point source subtraction goes down with increasing $k_{\parallel}$.

\begin{figure*}
\psfrag{kpara in Mpc-1}[lc][lc][1.5]{$k_{\parallel}$ [Mpc$^{-1}$]}\psfrag{kper in Mpc-1}[b][B][1.5]{$k_{\perp}$ [Mpc$^{-1}$]}\psfrag{K2 Mpc3}[c][t][1.5]{$P(k_{\perp},k_{\parallel})$ [K$^{2}$ Mpc$^{3}$]}
\psfrag{P10/Pf}[b][t][1.5]{$\kappa_{f}=P_{10}/P_{f}$}
\psfrag{ 0.01}[c][r][1.5]{$0.01$}\psfrag{ 0.1}[c][r][1.5]{$0.1$}\psfrag{ 1}[c][r][1.5]{$1$}
\psfrag{10}{$ $}
\psfrag{3}[c][l][1.5]{$10^{3}$}\psfrag{4}[c][l][1.5]{$10^{4}$}\psfrag{5}[c][l][1.5]{$10^{5}$}\psfrag{6}[c][l][1.5]{$10^{6}$}\psfrag{7}[c][l][1.5]{$10^{7}$}
\psfrag{ 0}[c][l][1.5]{$ $}\psfrag{ 10}[c][r][1.5]{$10$}\psfrag{ 20}[c][r][1.5]{$20$}\psfrag{ 30}[c][r][1.5]{$30$}\psfrag{ 40}[c][r][1.5]{$40$}\psfrag{ 50}[c][r][1.5]{$50$}\psfrag{ 60}[c][r][1.5]{$ $}\psfrag{ 5}[c][r][1.5]{$ $}\psfrag{ 15}[c][r][1.5]{$ $}\psfrag{ 25}[c][r][1.5]{$ $}\psfrag{ 35}[c][r][1.5]{$ $}\psfrag{ 45}[c][r][1.5]{$ $}
\psfrag{U}[cb][ct][1.0]{$\ell$}\psfrag{147}[cb][ct][1.5]{$147$}\psfrag{1470}[cb][lt][1.5]{$1470$}\psfrag{0.027}[c][c][1.5]{$ $}\psfrag{0.27}[c][c][1.5]{$0.27$}\psfrag{2.7}[c][c][1.5]{$2.7$}\psfrag{Tau}[cb][rt][1.0]{$\tau$ [ $\mu$s ]}
\includegraphics[width=170mm]{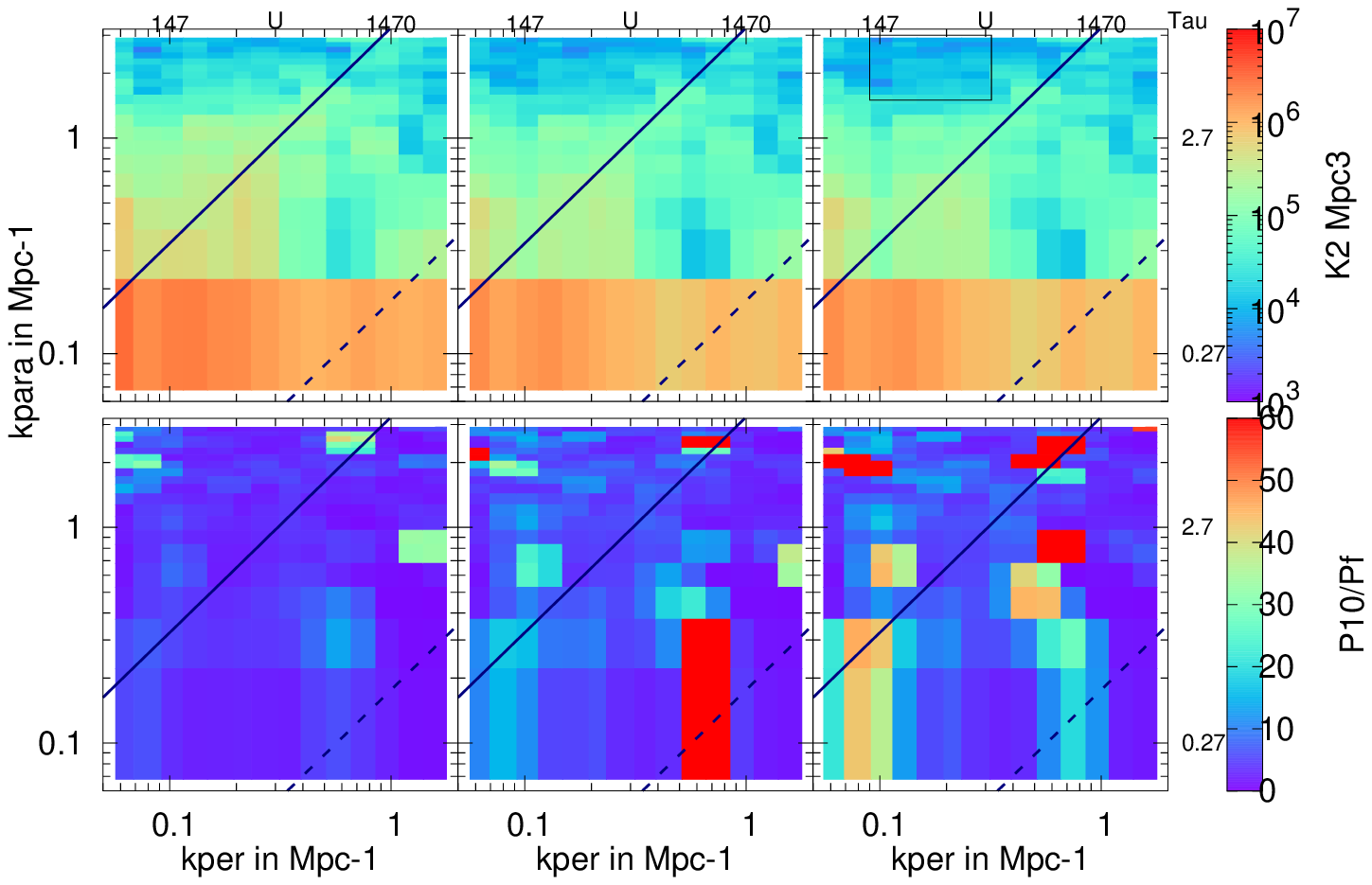}\vspace{8mm}
\caption{The upper row show the absolute value of the estimated cylindrical-binned power spectrum $P(k_{\perp},k_{\parallel})$ after point source subtraction for different tapering  $f=2.0,0.8,0.6$ (left to right panels). The lower row show  the corresponding $\kappa_{f}$ values.  In all the cases, the solid and dashed  lines respectively denote  $[k_{\parallel}]_{H}$
and $[k_{\parallel}]_{\vtha}$. Note that the ($k_{\perp},k_{\parallel}$) modes enclosed within the rectangular area indicated in the upper right panel at $f=0.6$, have been binned in the later part of the section to obtain the spherically binned averaged power spectrum $P(k)$.}
\label{pk_res}
\end{figure*}

The analysis  till now has been restricted to $f=10$ which essentially corresponds to no tapering. We now study  the effect of tapering by considering smaller values of $f$. This restricts the sky response by introducing a tapering function which falls before the first null of the PB. The tapering function gets narrower as $f$ is reduced. Further, we have already seen that point source subtraction considerably reduces the foreground power in some of the $(k_{\perp},k_{\parallel}) $ bins,  and we focus on the results after point source subtraction for the subsequent analysis.  
Figure \ref{pk_res} shows $\mid P(k_{\perp},k_{\parallel}) \mid$ (upper panels) for  $f=2.0,\,0.8\,\textrm{and}\,0.6$ from left to right respectively. We see that for all values of $f$ the overall structure is very similar to that for $f=10$  in Figure \ref{pk_10} with particularly large values of the power at $k_{\parallel} \le 0.2 \, \rm{Mpc^{-1}}$. At small $k_{\perp}$ the power falls considerably outside the wedge with values $\sim 10^{3}-10^{4}  \, \rm{K^2  \, Mpc^3}$, whereas at $ k_{\perp} \ge 0.5 \, \rm{Mpc^{-1}}$ the power falls to this range  at large $k_{\parallel}$ even inside the wedge.  Moreover we note that the values of the foreground power fall as the tapering is increased ($f$ is reduced). This change is pronounced between $f=2$ and $0.8$ and even between $f=0.8$ and $f=0.6$. In order to highlight the suppression of foreground power due to tapering we consider $\kappa_{f}=\mid [P(k_{\perp},k_{\parallel})]_{f=10.0}/[P(k_{\perp},k_{\parallel})]_f \mid$,  
 the ratio of the power with no tapering $(f=10)$ to that with tapering value $f$, shown in the lower panels of Figure \ref{pk_res}.  We find that the values of $\kappa_{f}$ are in the range 
$6.3 \times 10^{-3} - 1.2 \times 10^2$ for $f=2.0$ and the range changes to $ 8.4 \times 10^{-3} - 1.2 \times 10^{4} $ for $f=0.6$. However, the very small  values ($\kappa_{f} < 1 $) and the very large  values ($\kappa_{f}> 100$)  occur at only a few $ (k_{\perp},k_{\parallel})$ bins.
 In order to highlight  
 the overall variations of $\kappa_f$, we have restricted the   dynamical range in the lower panels  to  $1-60$ and shown the interpolated $\kappa_f$ values. Along the boundary of the  foreground wedge $\kappa_{f}$ is found to have values  approximately $< 40,\, 220\, {\rm and}\, 1200$ for $f = 2.0\,, 0.8\, \textrm{and}\, 0.6$ respectively. Overall we see that tapering is more effective at large $k_{\parallel}$, and  the large values of  $\kappa_f$ 
  are mainly located in the EoR window outside the foreground wedge. The prevalence of large $\kappa_f$ values also increases as tapering is increased. We also find that with the exception of the smallest $k_{\perp}$ bin, foreground suppression in the EoR window improves by around  a factor of $4$, if not more, when  $f$ is varied from $10$ to $0.6$.  

\begin{figure*}
\psfrag{1}[c][c][1.4]{$1$}
\includegraphics[width=140mm]{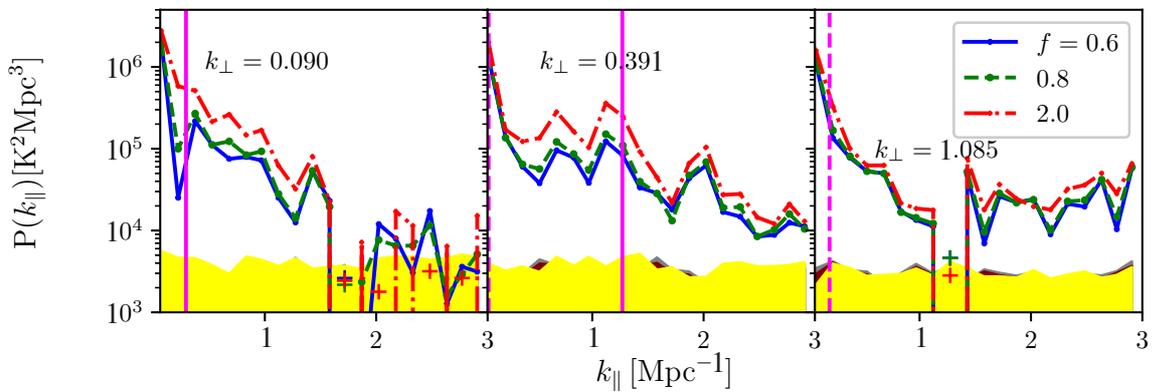}
\caption{The estimated cylindrical-binned power spectrum $P(k_{\perp},k_{\parallel})$ 
as a function of $k_{\parallel}$ at three representative values of $k_{\perp}$
after  point source subtraction for  $f=2.0,\,0.8$ and $0.6$. The $k_{\parallel}$ bins corresponding to the negative power values are indicated by blue, green and red `+' markers for $f=0.6,\,0.8\,\textrm{and}\,2.0$ respectively. The shaded regions show the $1\,\sigma$ errors $\delta{\rm  P}$ due to the system noise after point source subtraction at $f=2.0\,\textrm{(shown in yellow)}\,,\,0.8\,\textrm{(shown in maroon)}\, \textrm{and}\, 0.6\,\textrm{(shown in gray)}$. In all the cases the solid  and dashed pink vertical lines respectively show  $[k_{\parallel}]_{H}$
and $[k_{\parallel}]_{\vtha}$.}
\label{pk_all_slice}
\end{figure*}

We would now like to analyse the roll-off of the foreground contribution as $k_{\parallel}$ is increased. In particular, we would like to see how this is affected if we increase the  tapering. 
Figure \ref{pk_all_slice} shows  $P(k_{\perp}, k_{\parallel})$ as a function of $k_{\parallel}$ for three representative values of $k_{\perp}$ (same as  those shown in Figure \ref{pk_10_slice}) for  the three different tapering ($f=2.0,0.8$ and $0.6$). The $k_{\parallel}$ values corresponding to the negative power are indicated by `+' in Figure \ref{pk_all_slice} for all the values of `$f$'. For reference, the $1\,\sigma$ noise level ($\delta P_{N}$) for $f=2.0\,\textrm{(in yellow)},0.8\,\textrm{(in maroon)}\textrm{and}\,0.6\,\textrm{(in gray)}$ are also shown as shaded regions. We see that for all the values of $f$ the value of $P(k_{\perp},k_{\parallel})$ falls several orders of magnitude as $k_{\parallel}$ increases beyond the lowest $k_{\parallel}$ bin all the way to $k_{\parallel} \approx 1.5 \, \textrm{Mpc}^{-1}$ beyond which it oscillates with both positive and negative values which are a few times the $1\,\sigma$ noise level at both $k_{\perp}=0.090\, {\rm and} \, 1.085 \,{\rm Mpc}^{-1}$. The negative $P(k_{\perp}, k_{\parallel})$ values are found to be consistent with the $1\,\sigma$ noise level. This behaviour is quite similar to that seen earlier in Figure \ref{pk_10_slice} for $f=10$.
We see that the  foregrounds drop by  two to three 
orders of magnitude from the smallest to the largest $k_{\parallel}$ 
bins probed in our observation. In all cases the foregrounds are considerably smaller than those for $f=10$ 
(Figure \ref{pk_10_slice}). We notice that the  foreground roll-over gets steeper as we increase the tapering. 
The difference is particularly pronounced  between $f=2$ and $f=0.8$, the difference between $f=0.8$ and $0.6$ is small but still noticeable. At $k_{\perp}=0.090\,\textrm{Mpc}^{-1}$, we find $\sim 136$ times foreground  suppression near the horizon for  $f=0.6$ with respect to $f=2.0$ .  A somewhat smaller, but substantial,  foreground suppression is also noticed at $k_{\perp} = 0.391 \, \textrm{Mpc}^{-1}$. In contrast, the foreground power does not appear to fall much at $k_{\perp}=1.085\,\textrm{Mpc}^{-1}$ when $f$ is varied from $2.0$ to $0.6$. It may be noted that the power in this $k_{\perp}$ mode is rather low  around $k_{\parallel} \approx 0.4\,\textrm{Mpc}^{-1}$ even for $f=10$  (Figure \ref{pk_10_slice}) indicating that this may have a substantial  noise contribution. The difference between $f=2$ and the smaller $f$ values is relatively small for the $k_{\parallel}$ bins which are noise dominated.  As mentioned earlier, the effectiveness of  tapering is dependent on the baseline distribution. Tapering is expected to be more effective in the regions of baseline space ($k_{\perp}$) where we have a dense $uv$ sampling. 
  
\begin{figure}
\begin{center}
\includegraphics[width=85mm]{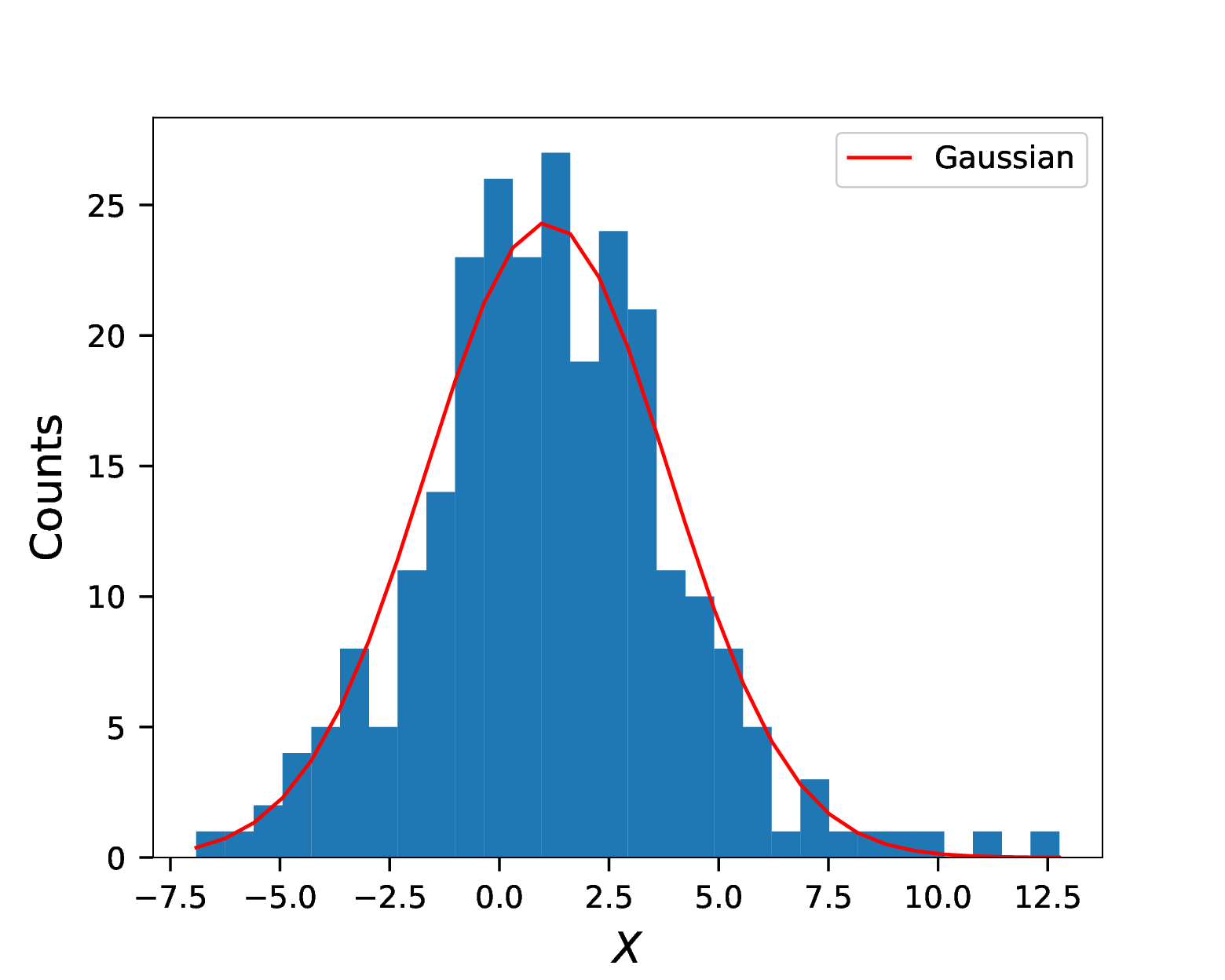}
\caption{The histogram of the variable $X=\frac{P(k_{\perp},\,k_{\parallel})}{\delta P_{N}(k_{\perp},\,k_{\parallel})}$. The red line shows the Gaussian fit with mean $1.1$ and standard deviation $2.77$.}
\label{pkbysigma}
\end{center}
\end{figure}

For the subsequent analysis we focus on the data after point source subtraction with $f=0.6$. We have also considered $f$ values smaller than $f=0.6$ (not shown here), however foreground suppression saturates around $f=0.6$ and does not improve much for smaller values of $f$. We have seen that in several $(k_{\perp}, k_{\parallel})$ bins  the estimated power spectrum $P(k_{\perp}, k_{\parallel})$ is comparable to the estimated r.m.s. fluctuation  $\delta P_{N}$   arising from system noise. We now identify a region outside the foreground wedge which has the least contribution from foreground leakage. We have selected a rectangular region  bounded by $0.09\,\textrm{Mpc}^{-1}\, \leq k_{\perp} \leq \,0.32\,\textrm{Mpc}^{-1}$ and  $1.5\,\textrm{Mpc}^{-1}\, \leq k_{\parallel} \leq \,3.0\,\textrm{Mpc}^{-1}$ which is  shown in the upper right panel of Figure \ref{pk_res}.  A visual inspection of the data reveals  the presence of relatively large foreground leakage in the two smallest $k_\perp$ bins, and we  have excluded these.  We use the quantity 
\begin{equation}
    X=\frac{P(k_{\perp},\,k_{\parallel})}{\delta P_{N}(k_{\perp},\,k_{\parallel})}
    \label{eq:x1}
\end{equation}
to quantify the statistics of the measured $P(k_{\perp},\,k_{\parallel})$ values within the rectangular region defined above. As mentioned earlier,  $\delta P_{N}(k_{\perp},\,k_{\parallel})$ is the predicted standard deviation arising from system noise alone. In the situation where there is no foreground contribution and  the estimated $P(k_{\perp},\,k_{\parallel})$ is entirely due to statistical  fluctuations arising from the system noise, we expect $X$ to have a Gaussian distribution with ${\rm mean}(X)=0$ and ${\rm var}(X)=1$.
 Figure \ref{pkbysigma} shows a histogram of $X$ where we see that the bulk of the data  may be described by a Gaussian distribution with ${\rm mean}(X)=1.1$ and $\sqrt{{\rm var}(X)}=2.77$. The positive mean indicates that we have residual foregrounds  still present within the rectangular region. The fact that we have ${\rm var}(X)>1$ indicates that 
 $\delta P_{N}(k_{\perp},\,k_{\parallel})$ underestimates  the actual statistical fluctuations in the measured  $P(k_{\perp},\,k_{\parallel})$ values, and the actual statistical errors $\delta P(k_{\perp},\,k_{\parallel})$ are a factor  $\sqrt{{\rm var}(X)}=2.77$ times larger than $\delta P_{N}(k_{\perp},\,k_{\parallel})$ {\it i.e.}  $\delta P(k_{\perp},\,k_{\parallel})= \sqrt{{\rm var}(X)} \times  \delta P_{N}(k_{\perp},\,k_{\parallel}) $. 
 The value  $[{\rm mean}(X) \, \times \, \delta P_{N}(k_{\perp},\,k_{\parallel})]$ provides an  estimate of the foreground contribution in the individual $P(k_{\perp},\,k_{\parallel}) $ measurements.  We note that the  level of  foreground leakage in the individual $P(k_{\perp},\,k_{\parallel}) $ measurements are much smaller than the estimated statistical fluctuations  $(P(k_{\perp},\,k_{\parallel}) \approx 0.4  \, \delta P(k_{\perp},\,k_{\parallel}))$, and these may be used to constrain the EOR 21-cm power spectrum.

\begin{figure}
\psfrag{k}[c][c][1.0]{$k$ [Mpc$^{-1}$]}\psfrag{Delta(k)K2}[c][l][1.0]{$\Delta^2(k)$ [K$^{2}$]}
\psfrag{ 1}[c][c][1.0]{$1$}
\psfrag{ 1.5}[c][c][1.0]{$1.5$}\psfrag{ 3}[c][c][1.0]{$3$}
\psfrag{m}{$\times10^{4}$}
\psfrag{ 10000}[c][c][1.0]{$1$}\psfrag{ 20000}[c][c][1.0]{$2$}\psfrag{ 30000}[c][c][1.0]{$3$}\psfrag{ 40000}[c][c][1.0]{$4$}\psfrag{-5000}[c][c][1.0]{$-0.5$}\psfrag{ 5000}[c][c][1.0]{$0.5$}\psfrag{ 15000}[c][c][1.0]{$1.5$}
\psfrag{ 25000}[c][c][1.0]{$2.5$}\psfrag{ 35000}[c][c][1.0]{$3.5$}
\psfrag{ 0}[c][c][1.0]{$0$}
\includegraphics[scale=0.6]{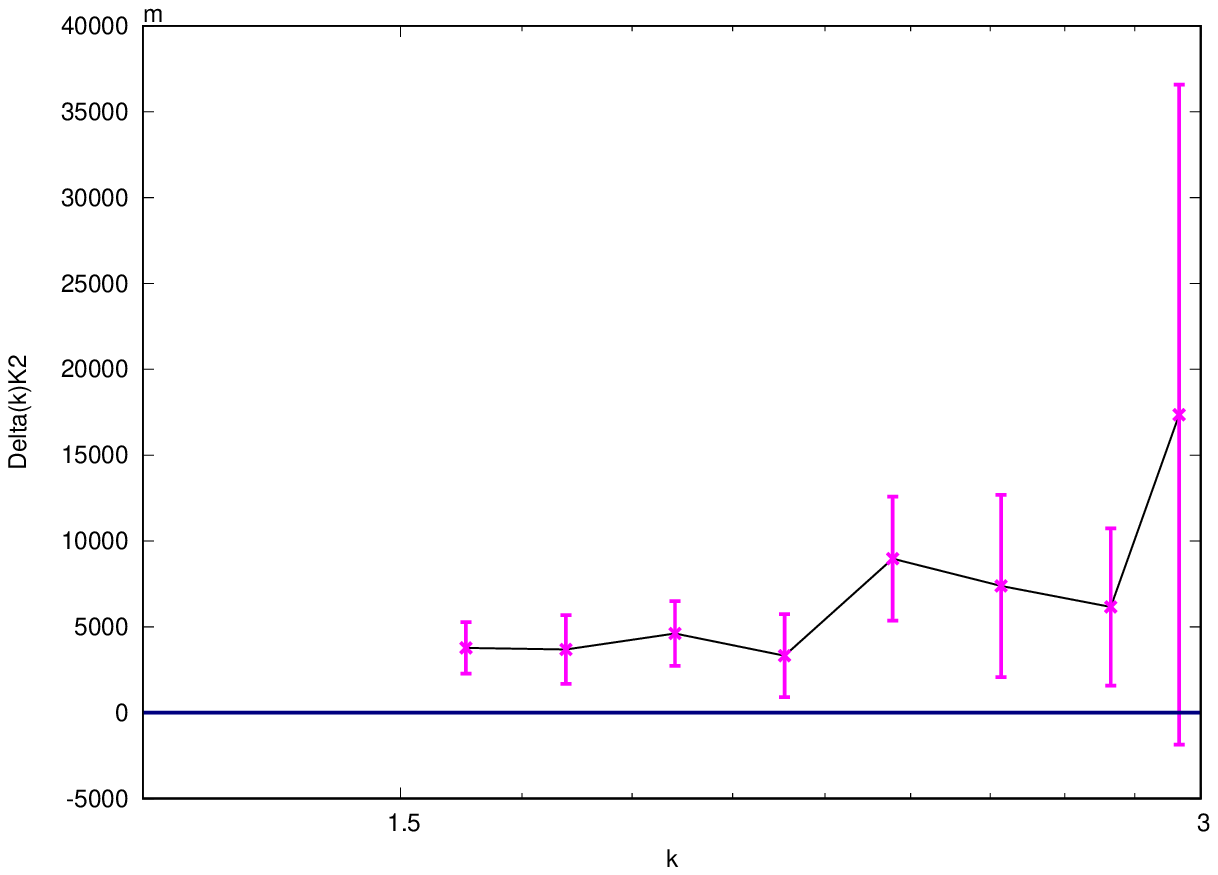}
\caption{The mean square brightness temperature fluctuations $\Delta^2(k)$
shown  as a function of $k$   along with $2 \,\sigma$ error bars.
}

\label{fin}
\end{figure}

We have spherically binned the $\kv$ modes within the rectangular region  in order to reduce the statistical fluctuations in the measured spherically binned power spectrum $P(k)$.
Figure \ref{fin} shows the  mean square brightness temperature fluctuations  $\Delta^{2}(k)= {k^{3}}P(k)/{2\pi^{2}} $ as a function of $k$ along with the $2\,\sigma$ error bars,  here $\sigma= {k^{3}}[\delta P(k)]/{2\pi^{2}} $. 
We summarize the results in Table \ref{Pk}  where the first three  columns respectively show $k$, $\Delta^{2}(k)$ and  $\sigma$ for each spherical bin. The estimated $\Delta^{2}(k)$ may be interpreted as arising from a combination of residual foregrounds plus  statistical fluctuations.  
 The fourth column of Table \ref{Pk} lists the $2\,\sigma$ upper limits on $\Delta^{2}(k)$ ($\Delta^{2}_{UL}(k)=\Delta^{2}(k)+2\,\sigma$; \citealt{mertens20}) corresponding to each $k$-bin. We find that we have the tightest constraint at $k=1.59\,\textrm{Mpc}^{-1}$  where we obtain the $2 \, \sigma $ upper limits of  $(72.66)^{2}$ K$^{2}$ on the mean squared HI 21-cm brightness temperature fluctuations.

\begin{table}
\caption{Estimated spherically binned mean square brightness temperature fluctuations $\Delta^2(k)$ and statistical error predictions $\sigma$ for the same. The $2\,\sigma$ upper limits on $\Delta^{2}(k)$ ($\Delta^{2}_{UL}(k)=\Delta^{2}(k)+2\,\sigma$) are listed corresponding to each $k$-bin.} 
\label{Pk}
\begin{tabular}{|l|c|c|c|c|}
\hline
\hline
$k\, \textrm{Mpc}^{-1}$ & $\Delta^2(k)\,\textrm{K}^2$ & $\sigma\,\textrm{K}^2$ &   Upper limit, $\Delta_{UL}^{2}(k)$ \\
 & $={k^{3}}P(k)/{2\pi^{2}}$ & $={k^{3}}\delta P/{2\pi^{2}}$ & $(\textrm{K})^2 \, [2\sigma]$ \,   \\
\hline
$1.59$ &	$(61.47)^2$ &	$(27.40)^2$ &	$(72.66)^2$ \\
\hline
$1.73$ &	$(60.70)^2$ &	$(31.61)^2$ &	$(75.38)^2$ \\
\hline
$1.90$ &	$(67.96)^2$ &	$(30.74)^2$ &	$(80.68)^2$ \\
\hline
$2.09$ &	$(57.61)^2$ &	$(34.75)^2$ &	$(75.72)^2$ \\
\hline
$2.30$ &	$(94.74)^2$ &	$(42.47)^2$ &	$(112.17)^2$ \\
\hline
$2.52$ &	$(85.93)^2$ &	$(51.53)^2$ &	$(112.67)^2$ \\
\hline
$2.78$ &	$(78.50)^2$ &	$(47.85)^2$ &	$(103.64)^2$ \\
\hline
$2.94$ &	$(131.75)^2$ &	$(98.00)^2$ &	$(191.22)^2$ \\
\hline
\hline
\end{tabular}
\end{table}

\section{Summary and Conclusions}

We have validated and demonstrated the capabilities  of the  TGE to measure the MAPS $C_{\ell}(\Delta \nu)$ and the 3D power spectrum $P(k_{\perp}, k_{\parallel})$ by applying it to a small data set from $150 \, {\rm MHz}$ GMRT observations. More than $47 \%$ of the data is flagged to avoid RFI and other systematic errors. We have carried out simulations (Section \ref{sim}) to verify that in the absence of foregrounds and noise our estimator is able to faithfully recover the input model power spectrum from simulated data which has exactly the same baseline distribution and flagging as the actual data. However, the present data is foreground dominated and further the system noise is much too large for a 21-cm signal detection. 
This  data has been analysed previously by \citet{ghosh3} who have used it to 
characterise the foregrounds at $150$ MHz. The earlier work has also modelled and subtracted out the point sources with flux $>9 \, {\rm mJy}$
from the central  $4.0^{\circ} \times 4.0^{\circ}$  region of the FoV. In this work we have analysed the data both before and after point source subtraction. The TGE offers a tapering parameter $f$ which can be varied to change the width of the Gaussian tapering window function whose FWHM is $f$ times the FWHM of the GMRT  PB. The value $f=10$ essentially corresponds to no tapering. The sky response of the tapering window function gets narrower as $f$ is reduced. We have considered $f=2,0.8$ and $0.6$ in our analysis. We find that  the effect of tapering saturates around $f=0.8$, and there is no further effect if $f$ is  reduced below $0.6$.  

Considering  the $\Delta \nu$ dependence of $C_{\ell}(\Delta\nu)$, for all values of $f$ we find that  in addition to a component which falls off smoothly with increasing $\Delta \nu$ we also have a component which oscillates with $\Delta \nu$. Before point source subtraction
 the amplitude of the oscillating component is around $3.0\% - 39.9 \%$ of the smooth component (modelled with a 3rd order polynomial in the range  $0.5 < \Delta\nu < 5.5 \, {\rm  MHz}$) for  $f=10.0$. The amplitude and period of the oscillations  both decrease with increasing values of $\ell$.  
 The amplitude of  the smooth component and the oscillations both  decrease after point source subtraction. 
 The amplitude of the oscillation also goes down if the value of $f$ is reduced, however some oscillations still persist even for  $f=0.6$.  These oscillatory features in the measured $C_{\ell}(\Delta \nu)$ pose a severe threat for measuring the 21-cm power spectrum. The foregrounds could be easily modelled and removed \citep{ghosh1,ghosh2} to separate out the 21-cm signal and noise if these oscillations were not present. We identify the dominant  oscillatory components  as arising from bright sources located between the first null of the PB and the horizon. Although tapering does suppress the contribution from such sources leading to a reduction in the amplitude of the oscillations, some oscillations persist even after tapering. Tapering, which is implemented through a convolution in the $uv$ plane, is sensitive to the baseline distribution. The GMRT $uv$ coverage is rather sparse and patchy, and the problem is further aggravated here by the severe flagging. This  explains why  oscillations with a reduced amplitude still persist after tapering. We expect tapering to be more effective in a situation where we have a denser and more uniform baseline distribution.

The Fourier transform relating  $P(k_{\perp}, k_{\parallel})$ to $C_{\ell}(\Delta \nu)$ has been implemented through a maximum likelihood estimator. For $f=10$ we have estimates of $P(k_{\perp}, k_{\parallel})$ in $15 \times 20$ bins in the range $0.05\,{\rm Mpc}^{-1} \le k_{\perp} \le 1.8 \,{\rm Mpc}^{-1}$ and  $0.07\,{\rm Mpc}^{-1} \le k_{\parallel} \le 2.9 \,{\rm Mpc}^{-1}$  respectively. The $(k_{\perp}, k_{\parallel})$ range changes slightly if $f$ is varied. 
We find that for all values of $f$ the foregrounds are largely contained within the foreground wedge, although there is also some leakage beyond the wedge boundary. Considering $f=10$ before point source subtraction 
(Figure~\ref{pk_10}) we see that the foregrounds are particularly large at $k_{\parallel} \le 0.2 \, {\rm Mpc^{-1}}$ which is within the foreground wedge, there is also a strong leakage beyond the wedge boundary in the range $0.09 \, \rm{Mpc^{-1}} < k_{\perp} < 0.5 \, \rm{Mpc^{-1}}$. There is an overall fall in the foreground power when the point sources are subtracted, however the overall structure of $P(k_{\perp}, k_{\parallel})$  is more or less unchanged. 
Considering the effect of tapering, we see that compared to $f=10$ the foreground power drops if $f$ is  reduced to $f=2,0.8$ and $0.6$ (Figure \ref{pk_res}).  The foreground suppression increases if $f$ is reduced. However the suppression saturates around $f=0.8$, and there is no effect if $f$ is reduced below $0.6$.  We find that foreground suppression in the EoR window improves by round a factor of $4$, if not more, when $f$ is varied from $10$ to $0.6$. Considering  several different cases,  Figures \ref{pk_10_slice} and \ref{pk_all_slice} show  $ P(k_{\perp},\,k_{\parallel})$ as a function of  $k_{\parallel}$ for representative values of $k_{\perp}$. We notice that the  foreground roll-over gets steeper as we decrease the value of $f$. In all cases the foreground power  falls $2$ to $3$  orders of magnitude as $k_{\parallel}$ increases beyond the lowest $k_{\parallel}$ bin all the way to $k_{\parallel} \approx 1.5 \, \textrm{Mpc}^{-1}$ beyond which it oscillates with both positive and negative values (observed at several $k_{\perp}$) which are a few times the $1\,\sigma$ noise level.

We find that the foreground leakage beyond the wedge boundary is considerably reduced when we taper the sky response. Considering $f=0.6$, we identify the region bounded by $0.09\,\textrm{Mpc}^{-1}\, \leq k_{\perp} \leq \,0.32\,\textrm{Mpc}^{-1}$ and $1.5\,\textrm{Mpc}^{-1}\, \leq k_{\parallel} \leq \,3.0\,\textrm{Mpc}^{-1}$ (Figure \ref{pk_res}) as least contaminated by foregrounds. We have used $X$ (eq.~\ref{eq:x1}) to analyze the statistics of the power spectrum measurements in this rectangular region. We find  (Figure \ref{pkbysigma}) that the bulk of the data  can be described by a Gaussian with ${\rm mean}(X)=1.1$ and   $\sqrt{{\rm var}(X)}=2.77$. Using this, we infer that  $\delta P_{N}(k_{\perp},\,k_{\parallel})$  (the system noise contribution only) underestimates $\delta P(k_{\perp},\,k_{\parallel})$, the actual statistical fluctuations   of the measured   $P(k_{\perp},\,k_{\parallel})$,   by a factor of $\sqrt{{\rm var}(X)}=2.77$. 
A variety of  factors including random calibration errors, residual point source contributions and man made  radio frequency interference can contribute to  the error budget causing it to exceed the value predicted from the system noise alone. We have accounted for this   by using  $\delta P(k_{\perp},\,k_{\parallel}) = 2.77 \times \delta P_{N}(k_{\perp},\,k_{\parallel})$ to estimate the statistical error of the measured $P(k_{\perp},\,k_{\parallel})$.  The positive mean indicates that we have residual foregrounds  still present within the rectangular region.  
We note that the  level of  foreground leakage in the individual $P(k_{\perp},\,k_{\parallel}) $ measurements are much smaller than the estimated statistical fluctuations  $(P(k_{\perp},\,k_{\parallel}) \approx 0.4  \, \delta P(k_{\perp},\,k_{\parallel}))$.
 We have spherically binned  the  $\kv$ modes within the rectangular region  region in order to reduce the statistical fluctuation in each bin. Table~\ref{Pk} lists the estimated dimensionless power spectra corresponding to the $7\,k$-bins along with the estimated statistical error and $2\,\sigma$ upper limit corresponding to each $k$-bin. We find that we have the tightest constraint at  $k = 1.59 \textrm{Mpc}^{-1}$, and we use this to place a $ 2 \sigma$  upper limit   
 of  $(72.66)^{2} \, \mathrm{K}^{2}$  on the mean squared HI 21-cm brightness temperature fluctuations.
 
The upper limits obtained here is rather large and is not of interest  to constrain  models of reionization. 
 We however note that the  aim of the present paper is somewhat different which is 
 to demonstrate the capabilities of a new estimator for the 21-cm power spectrum.  We find that the estimated power spectrum is consistent with the expected foreground and noise behaviour. 
 This demonstrates that our new estimator is able to correctly estimate the noise bias and subtracts this  out to yield an unbiased estimate of the power spectrum.
 We  establish that  the TGE effectively suppresses the foreground contribution by tapering the sky response at large angular separations from the phase center. Further, although more than $47 \%$ of the data is flagged, we find that  the estimated power spectrum does not exhibit any artifacts due to the missing frequency channels. We plan to apply this new power spectrum estimation technique to more sensitive observations in future. 
 
\section{DATA AVAILABILITY}
The data from this study are available upon reasonable request to the corresponding author.

\section{ACKNOWLEDGEMENTS}
We  thank the staff of GMRT for making this observation possible. GMRT is run by National Centre for Radio Astrophysics of the Tata Institute of Fundamental Research.The authors  would also like to thank the anonymous reviewer whose comments helped improve the manuscript. AG would like to acknowledge IUCAA, Pune for providing support through the associateship programme.  SB would like to acknowledge funding provided under the MATRICS grant SERB/F/9805/2019-2020 
of the Science \& Engineering Research Board, a statutory body of Department of Science \& Technology
(DST),Government of India.



\appendix


\bsp	
\label{lastpage}

\begin{thebibliography}{99}

\bibitem[\protect\citeauthoryear{Ali, Bharadwaj \& Chengalur}{2008}]{ali} Ali S. S., Bharadwaj S.,\& Chengalur J.~N., 2008, MNRAS, 385, 2166A
\bibitem[\protect\citeauthoryear{Asad, et al.}{2015}]{Asad15} Asad K.~M.~B., et al., 2015, MNRAS, 451, 3709
\bibitem[\protect\citeauthoryear{Asad, et al.}{2018}]{Asad18} Asad K.~M.~B., Koopmans L.~V.~E., Jeli{\'c} V., de Bruyn A.~G., Pandey V.~N., Gehlot B.~K., 2018, MNRAS, 476, 3051
\bibitem[Ali et al.(2015)]{zali15} Ali, Z.~S., Parsons, A.~R., Zheng, H., et al.\ 2015, \apj, 809, 61
\bibitem[\protect\citeauthoryear{Barry, et al.}{2019}]{Barry19} Barry N., et al., 2019, ApJ, 884, 1
\bibitem[\protect\citeauthoryear{Bernardi, et al.}{2009}]{Bern09} Bernardi G., et al., 2009, A \& A, 500, 965
\bibitem[\protect\citeauthoryear{Bharadwaj \& Sethi}{2001}]{BS01} Bharadwaj S., Sethi S.~K., 2001, JApA, 22, 293
\bibitem [\protect\citeauthoryear{Bharadwaj \& Ali }{2005}]{BA5} Bharadwaj S. , \& Ali S. S. 2005, \mnras, 356, 1519
\bibitem[\protect\citeauthoryear{Bharadwaj, et al.}{2019}]{Bh18} Bharadwaj S., Pal S., Choudhuri S., Dutta P., 2019, MNRAS, 483, 5694
\bibitem[\protect\citeauthoryear{Bowman et al.}{2009}]{bowman09} Bowman, J.~D., Morales, M.~F., \& Hewitt, J.~N.\ 2009, \apj, 695, 183
\bibitem[\protect\citeauthoryear{Chakraborty, et al.}{2019a}]{Cha1} Chakraborty A., et al., 2019, MNRAS, 487, 4102
\bibitem[\protect\citeauthoryear{Chakraborty, et al.}{2019b}]{Cha2} Chakraborty A., et al., 2019, MNRAS, 490, 243
\bibitem[Chapman et al.(2012)]{chapman12} Chapman, E., Abdalla, F.~B., Harker, G., et al.\ 2012, \mnras, 423, 2518 
\bibitem[\protect\citeauthoryear{Cheng, et al.}{2018}]{Cheng18} Cheng C., et al., 2018, ApJ, 868, 26
\bibitem[\protect\citeauthoryear{Choudhuri et al.}{2014}]{samir14} Choudhuri, S.,  Bharadwaj, S., Ghosh, A., \& Ali, S.~S.,\ 2014, \mnras, 445, 4351 
\bibitem[\protect\citeauthoryear{Choudhuri et al.}{2016a}]{samir16}  Choudhuri, S.,  Bharadwaj, S., Roy, N., Ghosh, A., \& Ali, S.~S.,\ 2016a, \mnras, 459, 151
\bibitem[\protect\citeauthoryear{Choudhuri et al.}{2016b}]{samir17} Choudhuri, S., Bharadwaj, S., Chatterjee, S., et al.,\ 2016b, \mnras, 463, 4093
\bibitem[\protect\citeauthoryear{Choudhuri, et al.}{2017}]{samir17a} Choudhuri S., Bharadwaj S., Ali S.~S., Roy N., Intema H.~T., Ghosh A., 2017, MNRAS, 470, L11
\bibitem[\protect\citeauthoryear{Choudhuri, Dutta \& Bharadwaj}{2019}]{ITGE} Choudhuri S., Dutta P., Bharadwaj S., 2019, MNRAS, 483, 3910
\bibitem[\protect\citeauthoryear{Choudhuri, et al.}{2020}]{samir20} Choudhuri S., Ghosh A., Roy N., Bharadwaj S., Intema H.~T., Ali S.~S., 2020, MNRAS, 494, 1936
\bibitem[\protect\citeauthoryear {Datta et al.}{2010}]{adatta10} Datta, A., Bowman, J. D., \& Carilli, C. L. 2010, \apj, 724, 526
\bibitem[\protect\citeauthoryear{Datta, Choudhury \& Bharadwaj}{2007}]{KD07} Datta K.~K., Choudhury T.~R., Bharadwaj S., 2007, MNRAS, 378, 119
\bibitem[\protect\citeauthoryear{DeBoer, et al.}{2017}]{DB17} DeBoer D.~R., et al., 2017, PASP, 129, 045001
\bibitem[\protect\citeauthoryear {Dillon et al.}{2014}]{dillon14} Dillon, J.~S., Liu, A., Williams, C.~L., et al.\ 2014, PRD, 89, 023002 
\bibitem[\protect\citeauthoryear {Dillon et al.}{2015}]{dillon15} Dillon, J.~S., Liu, A., Williams, C.~L., et al.\ 2015, PRD, 91(12), 123011
\bibitem [\protect\citeauthoryear{Di Matteo et al.}{2002}]{dmat1} Di Matteo, T., Perna, R., Abel, T. \& Rees, M.J., 2002, \apj, 564, 576
\bibitem[\protect\citeauthoryear{Eastwood, et al.}{2019}]{Eastwood19} Eastwood M.~W., et al., 2019, AJ, 158, 84
\bibitem[\protect\citeauthoryear {Furlanetto, Oh \& Briggs.}{2006}]{furla06} Furlanetto S. R., Oh S. P., Briggs F. H., 2006, Phys. Rep.,433, 181
\bibitem[\protect\citeauthoryear{Gehlot, et al.}{2019}]{Gehlot19} Gehlot B.~K., et al., 2019, MNRAS, 488, 4271
\bibitem[\protect\citeauthoryear{Ghosh et al.}{2011{\natexlab{a}}}]{ghosh1}  Ghosh, A., Bharadwaj, S., Ali, S.~S., \& Chengalur, J.~N.\ 2011{\natexlab{a}}, \mnras, 411, 2426
\bibitem[\protect\citeauthoryear{Ghosh et al.}{2011{\natexlab{b}}}]{ghosh2}  Ghosh, A., Bharadwaj, S.,  Ali, S.~S., \& Chengalur, J.~N.\ 2011{\natexlab{b}}, \mnras, 418, 2584
\bibitem[\protect\citeauthoryear{Ghosh, et al.}{2012}]{ghosh3} Ghosh A., Prasad J., Bharadwaj S., Ali S.~S., Chengalur J.~N., 2012, MNRAS, 426, 3295
\bibitem[\protect\citeauthoryear{Ghosh, et al.}{2020}]{Ghosh20} Ghosh A., et al., 2020, MNRAS, 495, 2813
\bibitem[\protect\citeauthoryear{Iacobelli, et al.}{2013}]{Ia13} Iacobelli M., et al., 2013, A\&A, 558, A72
\bibitem[\protect\citeauthoryear{Intema, et al.}{2017}]{Intema17} Intema H.~T., Jagannathan P., Mooley K.~P., Frail D.~A., 2017, A\&A, 598, A78
\bibitem[\protect\citeauthoryear{Jacobs, et al.}{2016}]{Jacobs16} Jacobs D.~C., et al., 2016, ApJ, 825, 114
\bibitem[Jeli{\'c} et al.(2008)]{jelic08} Jeli{\'c}, V., Zaroubi, S., Labropoulos, P., et al.\ 2008, \mnras, 389, 1319
\bibitem[\protect\citeauthoryear{Koopmans, et al.}{2015}]{Koo15} Koopmans L., et al., 2015, aska.conf,  1, aska.conf
\bibitem[\protect\citeauthoryear{Kolopanis, et al.}{2019}]{Kolopanis19} Kolopanis M., et al., 2019, ApJ, 883, 133
\bibitem[\protect\citeauthoryear{Li, et al.}{2019}]{Li19} Li W., et al., 2019, ApJ, 887, 141
\bibitem[\protect\citeauthoryear{Liu, Tegmark \& Zaldarriaga}{2009}]{Liu09} Liu A., Tegmark M., Zaldarriaga M., 2009, MNRAS, 394, 1575
\bibitem [\protect\citeauthoryear{Liu \& Tegmark}{2012}]{liu12} Liu, A., \& Tegmark, M.\ 2012, \mnras, 419, 3491
\textbf{}\bibitem [\protect\citeauthoryear{Liu et al.}{2014a}]{liu14a} Liu, A., Parsons, A.~R., \& Trott, C.~M.\ 2014a, PRD, 90, 023018
\bibitem [\protect\citeauthoryear{Liu et al.}{2014b}]{liu14b} Liu, A., Parsons, A.~R., \& Trott, C.~M.\ 2014b, PRD, 90, 023019
\bibitem[Liu et al.(2016)]{liu16} Liu, A., Zhang, Y., \& Parsons, A.~R.\ 2016, \apj, 833, 242
\bibitem[\protect\citeauthoryear{Mazumder, et al.}{2020}]{M20} Mazumder A., Chakraborty A., Datta A., Choudhuri S., Roy N., Wadadekar Y., Ishwara-Chandra C.~H., 2020, MNRAS, 495, 4071
\bibitem[\protect\citeauthoryear{McQuinn, et al.}{2006}]{mcquinn06} McQuinn M., Zahn O., Zaldarriaga M., Hernquist L., Furlanetto S.~R., 2006, ApJ, 653, 815
\bibitem[\protect\citeauthoryear{Mellema et al.}{2013}]{mellema13} Mellema, G., et al.\ 2013, Experimental Astronomy, 36, 235
\bibitem[\protect\citeauthoryear{Mertens, Ghosh \& Koopmans}{2018}]{mertens18} Mertens F.~G., Ghosh A., Koopmans L.~V.~E., 2018, MNRAS, 478, 3640
\bibitem[\protect\citeauthoryear{Mertens, et al.}{2020}]{mertens20} Mertens F.~G., et al., 2020, MNRAS, 493, 1662
\bibitem[\protect\citeauthoryear{Mondal, et al.}{2019}]{Mondal19} Mondal R., Bharadwaj S., Iliev I.~T., Datta K.~K., Majumdar S., Shaw A.~K., Sarkar A.~K., 2019, MNRAS, 483, L109
\bibitem[\protect\citeauthoryear{Mondal, et al.}{2020}]{Mondal20} Mondal R., et al., 2020, MNRAS, 494, 4043
\bibitem[\protect\citeauthoryear{Mondal, et al.}{2020}]{Mondal20a} Mondal R., et al., 2020, arXiv, arXiv:2004.00678
\bibitem[\protect\citeauthoryear{Morales \& Hewitt}{2004}]{Morales04} Morales M.~F., Hewitt J., 2004, ApJ, 615, 7
\bibitem[\protect\citeauthoryear{Morales}{2005}]{morales05} Morales M.~F., 2005, ApJ, 619, 678
\bibitem[\protect\citeauthoryear{Morales \& Matejek}{2009}]{morales09} Morales M.~F., Matejek M., 2009, MNRAS, 400, 1814
\bibitem[\protect\citeauthoryear{Morales \& Wyithe}{2010}]{morales10} Morales, M.~F., \& Wyithe, J.~S.~B.\ 2010, \araa, 48, 127
\bibitem[\protect\citeauthoryear{Paciga et al.}{2011}]{paciga11} Paciga G. et al., 2011, \mnras, 413, 1174
\bibitem[\protect\citeauthoryear{Paciga et al.}{2013}]{paciga13}  Paciga, G., Albert, J.~G., Bandura, K., et al.\ 2013, \mnras, 433, 639 
\bibitem[\protect\citeauthoryear{Parsons et al.}{2010}]{parsons10} Parsons A. R. et al., 2010, AJ, 139, 1468
\bibitem[Parsons et al.(2012)]{parsons12}  Parsons, A.~R., Pober, J.~C., Aguirre, J.~E., et al.\ 2012, \apj, 756, 165 
\bibitem[\protect\citeauthoryear{Patil, et al.}{2017}]{patil17} Patil A.~H., et al., 2017, ApJ, 838, 65
\bibitem[\protect\citeauthoryear{Patwa \& Sethi}{2019}]{Patwa19} Patwa A.~K., Sethi S., 2019, ApJ, 887, 52
\bibitem[\protect\citeauthoryear{Pen, et al.}{2009}]{pen09} Pen U.-L., et al., 2009, MNRAS, 399, 181
\bibitem[\protect\citeauthoryear{Planck Collaboration, et al.}{2016}]{PLANCK16b} Planck Collaboration, et al., 2016, A\&A, 596, A108
\bibitem[\protect\citeauthoryear{Pober et al.}{2013}]{pober13} Pober J.~C. et al., 2013, \apjl, 768, L36
\bibitem[Pober et al.(2014)]{pober14} Pober, J.~C., Liu, A.,Dillon, J.~S., et al.\ 2014, \apj, 782, 66
\bibitem[Pober et al.(2016)]{pober16} Pober, J.~C., Hazelton, B.~J., Beardsley, A.~P., et al.\ 2016, arXiv:1601.06177
\bibitem[\protect\citeauthoryear{Prasad \& Chengalur}{2012}]{jayanti} Prasad, J., \& Chengalur, J.\ 2012, Experimental Astronomy, 33, 157 
\bibitem[\protect\citeauthoryear{Prichard \& Loeb}{2012}]{prichard12} Pritchard, J. R. and Loeb, A., 2012, Reports on Progress in Physics 75(8), 086901
\bibitem[\protect\citeauthoryear{Saha, et al.}{2019}]{Saha19} Saha P., Bharadwaj S., Roy N., Choudhuri S., Chattopadhyay D., 2019, MNRAS, 489, 5866
\bibitem [\protect\citeauthoryear{Santos et al.}{2005}]{santos05} Santos, M.G., Cooray, A. \& Knox, L. 2005, 625, 575 
\bibitem[\protect\citeauthoryear{Shaver et al.}{1999}]{shaver99} Shaver, P.~A., Windhorst, R.~A., Madau, P., \& de Bruyn, A.~G.\ 1999, \aap, 345, 380
\bibitem[\protect\citeauthoryear{Shaw, et al.}{2014}]{Shaw14} Shaw J.~R., Sigurdson K., Pen U.-L., Stebbins A., Sitwell M., 2014, ApJ, 781, 57
\bibitem[\protect\citeauthoryear{Shaw, et al.}{2015}]{Shaw15} Shaw J.~R., Sigurdson K., Sitwell M., Stebbins A., Pen U.-L., 2015, PhRvD, 91, 083514
\bibitem[\protect\citeauthoryear{Swarup, et al.}{1991}]{swarup91} Swarup G., Ananthakrishnan S., Kapahi V.~K., Rao A.~P., Subrahmanya C.~R., Kulkarni V.~K., 1991, CuSc, 60, 95
\bibitem[\protect\citeauthoryear{Thyagarajan et al.}{2013}]{thyag13} Thyagarajan, N., Udaya Shankar, N., Subrahmanyan, R., et al.\ 2013, \apj, 776, 6 
\bibitem[Thyagarajan et al.(2015)]{thyag15} Thyagarajan, N., Jacobs, D.~C., Bowman, J.~D., et al.\ 2015, \apjl, 807, L28
\bibitem[\protect\citeauthoryear{Thyagarajan, Carilli \& Nikolic}{2018}]{Thyagarajan18} Thyagarajan N., Carilli C.~L., Nikolic B., 2018, PhRvL, 120, 251301
\bibitem[\protect\citeauthoryear{Thyagarajan, et al.}{2020}]{Thyagarajan20} Thyagarajan N., et al., 2020, arXiv, arXiv:2005.10275
\bibitem[\protect\citeauthoryear{Tingay et al.}{2013}]{tingay13} Tingay, S. et al. 2013, Publications of the Astronomical Society of Australia, 30, 7
\bibitem[\protect\citeauthoryear{Trott et al.}{2012}]{trott1} Trott, C.~M., Wayth, R.~B., \& Tingay, S.~J.\ 2012, \apj, 757, 101 
\bibitem[Trott et al.(2016)]{trott16} Trott, C.~M., Pindor, B., Procopio, P., et al.\ 2016, \apj, 818, 139 
\bibitem[\protect\citeauthoryear{Trott, et al.}{2020}]{Trott20} Trott C.~M., et al., 2020, MNRAS.tmp, doi:10.1093/mnras/staa414
\bibitem[\protect\citeauthoryear{var Haarlem et al.}{2013}]{haarlem} van Haarlem, M.~P., Wise, M.~W., Gunst, A.~W., et al.\ 2013, \aap, 556, A2 
\bibitem[\protect\citeauthoryear{Vedantham et al.}{2012}]{vedantham12} Vedantham, H., Udaya Shankar, N., \& Subrahmanyan, R.\ 2012, \apj, 745, 176
\bibitem[\protect\citeauthoryear{Yatawatta et al.}{2013}]{yata13} Yatawatta, S. et al. 2013, Astronomy \& Astrophysics, 550, 136

\end{thebibliography}
\end{document}